\documentclass[reprint,superscriptaddress,amsmath,amssymb,aps,nofootinbib]{revtex4-2}
\usepackage{graphicx}% Include figure files
\usepackage{dcolumn}% Align table columns on decimal point
\usepackage{bm}% bold math
\usepackage[colorlinks,linkcolor=blue,citecolor=blue]{hyperref}% add hypertext capabilities

\begin{document}

\title{Semidilute Principle for Gels} 

\author{Naoyuki~Sakumichi}
\thanks{These authors contributed equally: N. Sakumichi, T. Yasuda}
\affiliation{Graduate School of Engineering, The University of Tokyo, 7-3-1 Hongo, Bunkyo-ku, Tokyo, Japan.}
\author{Takashi~Yasuda}
\thanks{These authors contributed equally: N. Sakumichi, T. Yasuda}
\affiliation{Graduate School of Engineering, The University of Tokyo, 7-3-1 Hongo, Bunkyo-ku, Tokyo, Japan.}
\author{Takamasa~Sakai}
\email[Correspondence should be addressed to N.~Sakumichi or T.~Sakai:\\]{sakumichi@gel.t.u-tokyo.ac.jp;\,
sakai@gel.t.u-tokyo.ac.jp}
\affiliation{Graduate School of Engineering, The University of Tokyo, 7-3-1 Hongo, Bunkyo-ku, Tokyo, Japan.}

\date{\today}

\begin{abstract}
Polymer gels such as jellies and soft contact lenses are soft solids consisting of three-dimensional polymer networks swollen with a large amount of solvent.
For approximately 80 years, the swelling of polymer gels has been described using the Flory--Huggins mean-field theory.
However, this theory is problematic when applied to polymer gels with large solvent contents owing to the significant fluctuations in polymer concentration.
In this study, we experimentally demonstrate the superiority of the semidilute scaling law over the mean-field theory for predicting the swelling of polymer gels.
Using the semidilute scaling law, we experimentally determine the universal critical exponent $\nu$ of the self-avoiding walk via swelling experiments on polymer gels.
The experimentally obtained value $\nu\simeq 0.589$ is consistent with the previously reported value of $\nu\simeq 0.588$, which was obtained by precise numerical calculations.
Furthermore, we theoretically derive and experimentally demonstrate a scaling law that governs the equilibrium concentrations.
This scaling law contradicts the predictions made by de Gennes' $c^{*}$ theorem.
A major deficiency of the $c^*$ theorem is that the network elasticity, which depends on the as-prepared state, is neglected.
These findings reveal that the semidilute scaling law is a fundamental principle for accurately predicting and controlling the equilibrium swelling of polymer gels.
\end{abstract}

\maketitle

\section{Introduction}
\label{sec:Introduction}

The swelling of polymer networks by absorption of surrounding solvent is a ubiquitous phenomenon \cite{flory1953principles,de1979scaling}; for example, it is used in water-absorbent materials in disposable diapers and portable toilets. 
Here, polymer networks swollen with a large amount of solvent are called polymer gels, such as jellies and soft contact lenses. 
As-prepared polymer gels after fabrication are generally not in an equilibrium swollen state and swell when immersed in a solvent. 
Revealing the governing law for the swelling of polymer gels has long been an important issue \cite{krishnamurti1929mechanism,flory1943jr}, and considerable studies have been conducted to predict the equilibrium swollen state (e.g., the liquid content in swollen polymer gels) \cite{flory1943jr,munch1977inelastic,beltzung1983investigation,bastide1984small,geissler1988compressional,horkay1989effects,rubinstein1996elastic,lang2014swelling,van2015biopolymer,obukhov1994network,urayama1996elastic,urayama1996crossover,candau1981experimental,patel1992elastic,neuburger1988critical,wu2004modeling,akalp2015determination,horkay2000osmotic}. 
This is because it is essential for a fundamental understanding of polymer gels and their applications. 
For example, in practical biomedical applications such as drug delivery \cite{peppas1997hydrogels}, adhesion barriers \cite{porchet1998clinical}, and artificial vitreous humor \cite{hayashi2017fast}, understanding the swelling behavior of polymer gels is essential because the liquid content significantly influences the softness and mass diffusivity of polymer gels in the external environment.\\

\begin{figure}[t!]
\centering
\includegraphics[width=\linewidth]{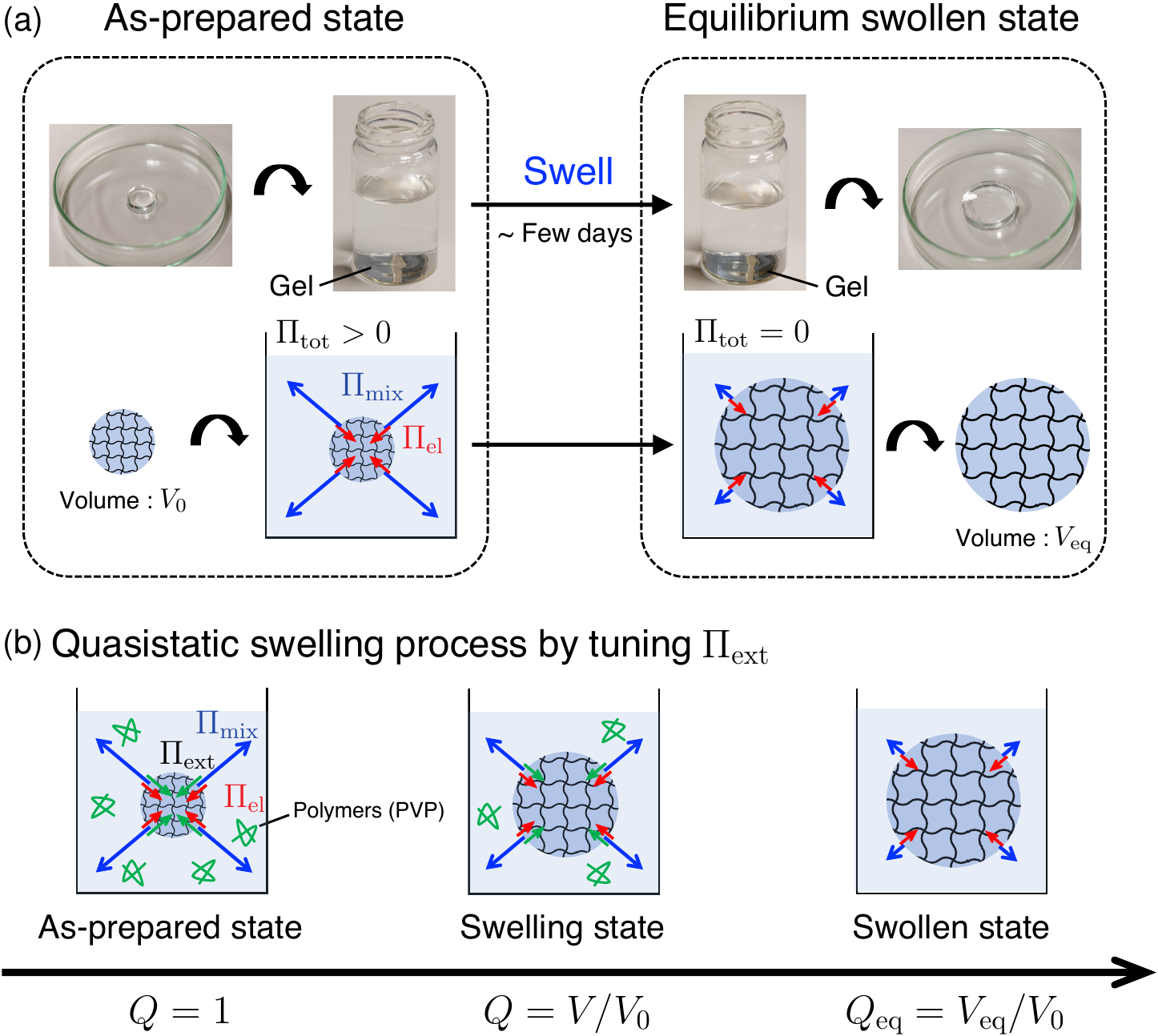}
\caption{
Swelling behavior of polymer gels.
(a) Volume change of a polymer gel from the as-prepared to equilibrium swollen states in a pure solvent.
A gel after fabrication is called the as-prepared state, where the volume is $V_{0}$.
Upon immersing into a pure solvent, the as-prepared gel swells because its osmotic pressure ($\Pi_\mathrm{tot}=\Pi_\mathrm{mix}+\Pi_\mathrm{el}>0$) is larger than the osmotic pressure outside ($\Pi_\mathrm{ext}=0$).
After a few days, the gel reaches the equilibrium swollen state at which $\Pi_\mathrm{tot}=0$, changing its volume ($V_{0} \to V_\mathrm{eq}$) and polymer mass concentration ($c_{0} \to c_\mathrm{eq}$).
(b) Quasistatic swelling process of a polymer gel by tuning the external osmotic pressure $\Pi_\mathrm{ext}$.
We tuned $\Pi_\mathrm{ext}$ using controlled aqueous poly(vinylpyrrolidone) (PVP) solutions, whose polymer mass concentration dependence of osmotic pressure was previously measured by Vink \cite{vink1971precision}.
At each equilibrium state, the osmotic pressure inside ($\Pi_\mathrm{tot}$) and outside ($\Pi_\mathrm{ext}$) a gel is equivalent: $\Pi_\mathrm{tot}=\Pi_\mathrm{mix}+\Pi_\mathrm{el}=\Pi_\mathrm{ext}$.
The volume change of a gel is quantified by the volume swelling ratio $Q\equiv V/V_{0}$, where $V$ is measured from a gel taken out from the solvent.
}
\label{fig:illustration}
\end{figure}

According to Flory and Rehner \cite{flory1943jr}, the swelling equilibrium of (electrically neutral) polymer gels in good solvents is determined by the balance between the elastic ($\Pi_\mathrm{el}$) and mixing ($\Pi_\mathrm{mix}$) contributions in the total swelling pressure of polymer gels [Fig.~\ref{fig:illustration}(a)]: 
\begin{equation}
\Pi_\mathrm{tot}=\Pi_\mathrm{mix}+\Pi_\mathrm{el}.
\label{eq:flory-rehner}
\end{equation}
The elastic contribution satisfies
$\Pi_\mathrm{el}=-G_{0}(c/c_{0})^{1/3}$, \cite{james1949simple,horkay2000osmotic}
because uniform swelling reduces the density of the polymer network.
Here, $G_{0}$ is the shear modulus in the as-prepared state, and $c_{0}$ and $c$ are the polymer mass concentrations in the as-prepared and equilibrium states, respectively.
Therefore, we consider the expression for $\Pi_\mathrm{mix}$.\\

A conventional approach to describe $\Pi_\mathrm{mix}$ in Eq.~(\ref{eq:flory-rehner}) is the Flory--Huggins (FH) mean-field theory, which was originally developed for polymer solutions \cite{flory1942thermodynamics,huggins1942some,huggins1942theory}.
The FH mean-field theory is applicable to concentrated polymer systems such as non-dilute polymer solutions and polymer blends with small concentration fluctuations \cite{patterson1967thermodynamics}, because it uses the mean-field approximation.
Thus, its application to dilute systems, including polymer gels with low polymer volume fractions (typically less than $0.1$), is problematic, as pointed out by Flory himself\footnote{
Regarding the limitations of the FH mean-field theory, Flory stated, ``Of importance here is the realization that the theory as developed so far is inappropriate, generally speaking, for dilute polymer solutions'' on p.~505 in Ref.~\cite{flory1953principles}.}.
However, in numerous subsequent studies, the FH theory has been inappropriately applied to polymer gels containing a large amount of solvent \cite{neuburger1988critical,patel1992elastic,wu2004modeling,akalp2015determination}, resulting in significant inconsistencies in the reported polymer--solvent interaction parameters \cite{merrill1993partitioning,pedersen2005temperature,venohr1998static,strazielle1968light}.\\

Another well-known claim, de Gennes' $c^{*}$ theorem \cite{de1979scaling}, asserts that the polymer concentration $c_\mathrm{eq}$ of a polymer gel in the equilibrium swollen state is proportional to the overlap concentration $c^{*}$ of a group of polymer chains in a good solvent.
Notably, this so-called $c^{*}$ ``theorem'' is not a mathematical theorem but only a physical conjecture that requires experimental validation.
This conjecture follows from the assumption that subchains (chains between adjacent crosslinks) in an equilibrium swollen network of a polymer gel disinterpenetrate, i.e., subchains do not interpenetrate each other and remain in contact at the overlap threshold.
Here, polymer gels are considered as a set of closely packed polymer chains sealed together by crosslinks, similar to polymer solutions at the overlap threshold $c^*$.
Several experimental observations have supported this conjecture, revealing similarities in the scaling relations between polymer gels and semidilute solutions \cite{munch1977inelastic,beltzung1983investigation,bastide1984small}.
However, subsequent systematic measurements of the elastic modulus of gels contradict the disinterpenetration of the swollen network chains assumed in this conjecture \cite{patel1992elastic,obukhov1994network,urayama1996elastic}.
Hence, the $c^*$ theorem, although seemingly plausible, remains controversial.\\

In this study, we measure $\Pi_\mathrm{mix}$ of chemically crosslinked polymer gels with precisely controlled homogeneous network structures \cite{sakai2008design} throughout the quasistatic swelling process from the as-prepared to equilibrium state [Fig.~\ref{fig:illustration}(b)], using the osmotic deswelling method proposed in 1940s \cite{boyer1945deswelling} and developed in the 1980s \cite{bastide1981osmotic}.
Consequently, we successfully demonstrate that the semidilute scaling law of polymer solutions \cite{de1979scaling,des1975lagrangian} governs $\Pi_\mathrm{mix}$ of polymer gels in Eq.~(\ref{eq:flory-rehner}).
Furthermore, by assuming the semidilute scaling law as the fundamental principle for polymer gels, we theoretically derive and experimentally validate the following three arguments.
First, the semidilute scaling law is superior to the FH mean-field theory in accurately predicting the equilibrium state of polymer gels throughout the quasistatic swelling process.
Second, swelling experiments of polymer gels provide a novel method for experimentally determining the excluded volume exponent $\nu$; we obtain $\nu\simeq 0.589$, which is consistent with those reported previously (see Table~\ref{tab:nu} below).
Here, $\nu$ is also known as the universal critical exponent of the self-avoiding walks (SAW), which corresponds to $n\to0$ in $O(n)$-symmetric university classes in three dimensions \cite{guida1998critical,pelissetto2002critical,clisby2016high,kompaniets2017minimally,shimada2016conformal}.
Third, we derive a novel scaling law for the equilibrium concentrations $c_\mathrm{eq}$ (i.e., the polymer mass concentrations in the equilibrium swollen states) that depends explicitly on the polymer mass concentration $c_0$ and shear modulus $G_0$ in the as-prepared state.
By contrast, $c_\mathrm{eq}$ predicted by the $c^{*}$ theorem does not depend on $c_0$ and $G_0$.
Our experimental results support the derived scaling law and contradict the prediction made by the $c^{*}$ theorem \cite{de1979scaling}.\\

The remainder of this paper is organized as follows. 
In Sec.~\ref{sec:Theory}, we formulate the FH mean-field theory and the semidilute scaling law of $\Pi_\mathrm{mix}$.
In Sec.~\ref{sec:Materials}, we explain the materials and methods. 
In Sec.~\ref{sec:Semidilute}, we analyze our results on swelling experiments by the FH theory and semidilute scaling law.
In Sec.~\ref{sec:prediction}, we compare the prediction of the quasistatic swelling process using the FH theory and semidilute scaling law.
In Sec.~\ref{sec:c* theorem}, we demonstrate the scaling law governing equilibrium concentration, which contradicts the predictions made by de Gennes' $c^{*}$ theorem.
In Sec.~\ref{sec:Conclusion}, we summarize the main results of this study. 
Several details are described in the Appendices to avoid digressing from the main subject.

\section{Flory--Huggins mean-field theory and semidilute scaling law of $\Pi_\mathrm{mix}$}
\label{sec:Theory}

To determine the law governing $\Pi_\mathrm{mix}$ of polymer gels, we experimentally evaluated $\Pi_\mathrm{mix}$ throughout the quasistatic swelling process, from the as-prepared to equilibrium swollen states, via osmotic deswelling \cite{boyer1945deswelling,bastide1981osmotic,yasuda2020universal}.
As shown in Fig.~\ref{fig:illustration}(b), we immerse a gel sample in an external polymer solution with a certain external osmotic pressure $\Pi_\mathrm{ext}$.
The equilibrium condition is $\Pi_\mathrm{tot}=\Pi_\mathrm{ext}$, at which the osmotic pressure inside and outside the gel is equivalent.
Especially, $\Pi_\mathrm{ext}=0$ for a pure solvent, as shown in Fig.~\ref{fig:illustration}(a).
Hence, by gradually tuning $\Pi_\mathrm{ext}$ with the controlled aqueous polymer solutions \cite{vink1971precision}, we can evaluate $\Pi_\mathrm{mix}$ as
\begin{equation}
\Pi_\mathrm{mix}=
\Pi_\mathrm{ext}
+G_{0}
\left(
\frac{c}{c_{0}}
\right)^{1/3},
\label{eq:Pos-gel}
\end{equation}
where $G_{0}$ and $c$ can be experimentally measured.
We assume Eq.~(\ref{eq:Pos-gel}) throughout this study.
In this study, we consider only the equilibrium states of the polymer gels in external polymer solutions, such as the as-prepared ($Q=1$ and $\Pi_\mathrm{ext}>0$), swelling ($Q>1$ and $\Pi_\mathrm{ext}>0$), and equilibrium swollen ($Q>1$ and $\Pi_\mathrm{ext}=0$) states.
Here, $Q\equiv V/V_{0}$ is the volume swelling ratio [see Fig.~\ref{fig:illustration}(b)].\\

In a solution with sufficiently long polymer chains, the Flory--Huggins (FH) mean-field theory \cite{flory1942thermodynamics,huggins1942some,huggins1942theory} predicts that
\begin{equation}
\Pi_\mathrm{mix}=
-\frac{k_BT}{v}
\left[
\phi+\ln\left(1-\phi\right)+\chi\phi^2
\right],
\label{eq:flory-huggins}
\end{equation}
where $k_B$, $T$, $v$, and $\phi$ are the Boltzmann constant, absolute temperature, volume of solvent molecules, and polymer volume fraction, respectively.
The FH interaction parameter $\chi$ quantifies the energies of the interactions between the polymer units and the solvent molecules per lattice site as follows:
$\chi \equiv z
\left(
\epsilon_\mathrm{pp}+\epsilon_\mathrm{ss}-2\epsilon_\mathrm{ps}
\right)
/(2k_{B}T)$, 
where $z$ is the coordination number of the lattice, and $\epsilon_\mathrm{pp}$, $\epsilon_\mathrm{ss}$, and $\epsilon_\mathrm{ps}$ are the polymer--polymer, solvent--solvent, and polymer--solvent interaction energies, respectively.
Although the FH theory is applicable only to concentrated polymer solutions \cite{patterson1967thermodynamics}, it has been inappropriately applied to polymer gels containing a large amount of solvent, by suitably adjusting $\chi$ \cite{neuburger1988critical,patel1992elastic,wu2004modeling,akalp2015determination}.
Consequently, significant inconsistencies in $\chi$ were reported in previous studies, although $\chi$ should be constant for the same polymer--solvent system and temperature \cite{bawendi1988systematic,pesci1989lattice}.
For example, $\chi=0.426$ for the poly(ethylene glycol) (PEG) hydrogel~\cite{merrill1993partitioning} and $\chi=0.366$, $0.441$, and $0.465$ for the aqueous PEG solution with the molar masses $M=4.6$~\cite{pedersen2005temperature}, $10$~\cite{venohr1998static}, and $33$~kg/mol~\cite{strazielle1968light}, respectively, were reported at $T=298$~K, using $v=18\times10^{-6}~\mathrm{m^{3}/mol}$.\\

By contrast, we propose that the semidilute scaling law of polymer solutions \cite{de1979scaling,des1975lagrangian} governs $\Pi_\mathrm{mix}$ in polymer gels, based on a recent observation~\cite{yasuda2020universal}.
In Ref.~\cite{yasuda2020universal}, it was observed that a universal osmotic equation of state of polymer solutions \cite{des1975lagrangian,de1979scaling,noda1981thermodynamic,ohta1982conformation,higo1983osmotic} describes $\Pi_\mathrm{mix}$ throughout the gelation process.
For polymer solutions, the semidilute scaling law is expressed as $\Pi_\mathrm{mix}M/(cN_{A}k_{B}T)=K(c/c^{*})^{1/(3\nu-1)}$, where $K\simeq 1.1$ is the universal constant for the semidilute polymer solutions, $N_A$ is the Avogadro constant, $c^{*}$ is the overlap concentration, and $\nu\simeq 0.588$ is the universal critical exponent of the SAW \cite{guida1998critical,pelissetto2002critical,clisby2016high,kompaniets2017minimally,shimada2016conformal}.
However, for polymer gels, $c^{*}\to 0$ and $c/c^{*}\to \infty$~\cite{yasuda2020universal}, because the molar mass of a polymer network is infinite.
Thus, we introduce the regulator $M_\mathrm{seg}$ to propose the following semidilute scaling law:
\begin{equation}
\frac{\Pi_\mathrm{mix}}{nk_BT}
=K\left(\frac{n}{n^{*}_\mathrm{seg}}\right)^{\frac{1}{3\nu-1}},
\label{eq:scaling-gel}
\end{equation}
as a fundamental principle for polymer networks that contain a large amount of solvent.
Here, both sides of Eq.~(\ref{eq:scaling-gel}) are dimensionless;
$n\equiv cN_{A}/M_\mathrm{seg}$ is the number density of a ``segment'' of polymers, where $M_\mathrm{seg}$ is the molar mass of the segment.
In this study, we use $M_\mathrm{seg}=44 \times 10^{-3}$ kg$/$mol, which is the molar mass of an ethylene glycol unit as a segment.
In Eq.~(\ref{eq:scaling-gel}), $n^{*}_\mathrm{seg}$ is a parameter that depends on $T$ and the type of polymer network and solvent considered.
We determined $n^{*}_\mathrm{seg}$ by adjusting to the experimental data.
This procedure is similar to the determination of $\chi$ in the FH mean-field theory from the experimental data.\\

Notably, a few pioneering studies \cite{bastide1984small,geissler1988compressional,horkay1989effects,rubinstein1996elastic,lang2014swelling,van2015biopolymer} attempted to describe the equilibrium swelling of polymer gels using the semidilute scaling law such as
$\Pi_\mathrm{mix}\sim\phi^{9/4}$.
However, they did not focus on demonstrating the superiority of the semidilute scaling law over the FH mean-field theory, owing to the limited accuracy of the gel experiments and the lack of theoretical comprehension of polymer gels.
By contrast, in this study, we experimentally demonstrate the superiority of Eq.~(\ref{eq:scaling-gel}) over the FH mean-field theory [Eq.~(\ref{eq:flory-huggins})] in terms of predicting the equilibrium state of polymer gels throughout the quasistatic swelling process.
We verify the superiority of the semidilute scaling law using precisely controlled homogeneous networks \cite{sakai2008design} and a different definition of the polymer mass concentration \cite{yasuda2020universal} that enables us to extend the universality of the osmotic equation of state of polymer solutions.

\section{Materials and methods}
\label{sec:Materials}

\subsection{Fabrication of model gels}
\label{sec:Fabrication}

As a model system to examine the quasistatic swelling process of chemically crosslinked polymer gels, we used a tetra-branched PEG hydrogel \cite{sakai2008design}, synthesized via the \textit{AB}-type cross-end coupling of two prepolymer (tetra-arm PEG) units of equal size. 
Each end of the tetra-arm PEG was modified with a mutually reactive maleimide (tetra-PEG MA) and thiol (tetra-PEG SH).
We dissolved tetra-PEG MA and tetra-PEG SH (NOF Co., Japan \& XIAMEN SINOPEG BIOTECH Co., Ltd., China) in phosphate citrate buffer with an ionic strength of $100$~mM and a \textit{p}H of $3.8$.
For gelation, we mixed these two solutions with equal molar masses $M$ for equal prepolymer mass concentrations $c_{0}$ at various mixing fractions $s$.
Here, $s$ is the molar fraction of the minor prepolymers (tetra-PEG SH) to all prepolymers ($0\leq s\leq 1/2$).
By tuning $s$, the desired connectivity $p$ can be obtained in accordance with $p = 2s$ \cite{sakai2016sol,yoshikawa2019connectivity}, where $p$ is defined as the fraction of reacted terminal functional groups to all terminal functional groups ($0\leq p \leq 1$).
Here, $p=1$ and $p<1$ correspond to the stoichiometrically balanced and imbalanced mixing of the prepolymer solutions, respectively.
Each mixed sample was maintained in an enclosed space to maintain humid conditions at room temperature ($T\simeq 298$ K) during the reaction.

\subsection{Swelling experiments}
\label{sec:Swelling experiments}

To investigate the quasistatic swelling (or deswelling) process of polymer gels from the as-prepared to equilibrium swollen states [Fig.~\ref{fig:illustration}(b)], we directly immerse a gel sample in an external aqueous polymer [poly(vinylpyrrolidone), PVP, K90, Sigma Aldrich] solution at a certain concentration $c_\mathrm{ext}$ at $T=298$~K with stirring \cite{boyer1945deswelling,bastide1981osmotic,yasuda2020universal}.
According to Vink \cite{vink1971precision}, the osmotic pressure $\Pi_\mathrm{ext}$ of aqueous polymer (PVP) solutions is given by
$\Pi_\mathrm{ext}\simeq
21.27c_\mathrm{ext}+
1.63{c_\mathrm{ext}}^{2}+
0.0166{c_\mathrm{ext}}^{3}$
for PVP mass concentration $c_\mathrm{ext}\leq200$ g$/$L.
By tuning $c_\mathrm{ext}$, we obtain the desired external osmotic pressure $\Pi_\mathrm{ext}$.
Here, $c_\mathrm{ext}$ is defined as the polymer (PVP) weight divided by the solution volume, and is different from the definition of the PEG mass concentrations ($c_{0}$, $c$, and $c_\mathrm{eq}$).\\

To experimentally determine $\Pi_\mathrm{mix}$ of a gel sample from Eq.~(\ref{eq:Pos-gel}), we measured the weight change ($W_{0} \to W$) of a gel sample from the as-prepared to equilibrium state using an electronic scale.
By calculating the swelling-induced (or deswelling-induced) change in the solvent volume from the weight change of a gel sample, we evaluate the polymer mass concentration ($c$), volume fraction ($\phi$), and volume swelling ratio ($Q=V/V_0$) at the equilibrium state.
Here, $V_{0}$ and $V$ are the volumes at the as-prepared and equilibrium states, respectively [Fig.~\ref{fig:illustration}(b)]. 
In addition, we measured the weight change of each gel sample once a day to determine that the equilibrium was reached.
The details of the evaluation and the reaching equilibrium are provided in Appendix~\ref{App:achievement}.\\

In particular, we evaluated $\Pi_\mathrm{mix}$ in the as-prepared state, by tuning $\Pi_\mathrm{ext}$ to maintain the volume ($Q=1$) and polymer mass concentration ($c=c_{0}$) of each gel sample.
Here, we determined $\Pi_\mathrm{ext}$, which yields $Q=1$ for each gel sample, by interpolating the $c_\mathrm{ext}$ dependence of $Q$ as a linear function around $Q=1$.
Further details are provided in Ref.~\cite{yasuda2020universal}.

\subsection{Measurement of shear modulus}
\label{sec:Measurement}

In this study, we used experimentally measured shear modulus $G_0$ in Eq.~(\ref{eq:Pos-gel}) without assuming any specific model, such as the affine and phantom network models.
This is to avoid the problem concerning the negative energy elasticity \cite{yoshikawa2021negative,Sakumichi2021,fujiyabu2021temperature,Shirai2022};
it has recently been found that $G_0=aT+b$ has a significantly large negative term $b$ in a narrow temperature ($T$) range.
This finding contradicts the conventional models that assume $G_0$ is approximately proportional to the absolute temperature $T$.\\

We measured $G_{0}$ of each as-prepared gel sample using dynamic viscoelasticity measurement.
Immediately after mixing the tetra-PEG MA and tetra-PEG SH aqueous solutions for $60$ seconds, the resulting solutions were poured into the gap of the double cylinder of a dynamic shear rheometer (MCR 301 and 302, Anton Paar, Graz, Austria).
Then, we cyclically sheared the samples in the gap between the inner cylinder and the outer cup at $298$~K for $6\times 10^4$ seconds, yielding the time course of the storage modulus $G'$.
The applied frequency and strain were $1$ Hz and $1$~\%, respectively.
After $G'$ reached equilibrium, which corresponded to the completion of the chemical reaction of the MA and SH functional groups, we determined $G'$ at $1$ Hz as the (equilibrium) shear modulus $G_{0}$.
We provide the experimental results in Appendix~\ref{App:shear modulus}.

\begin{figure}[t!]
\centering
\includegraphics[width=\linewidth]{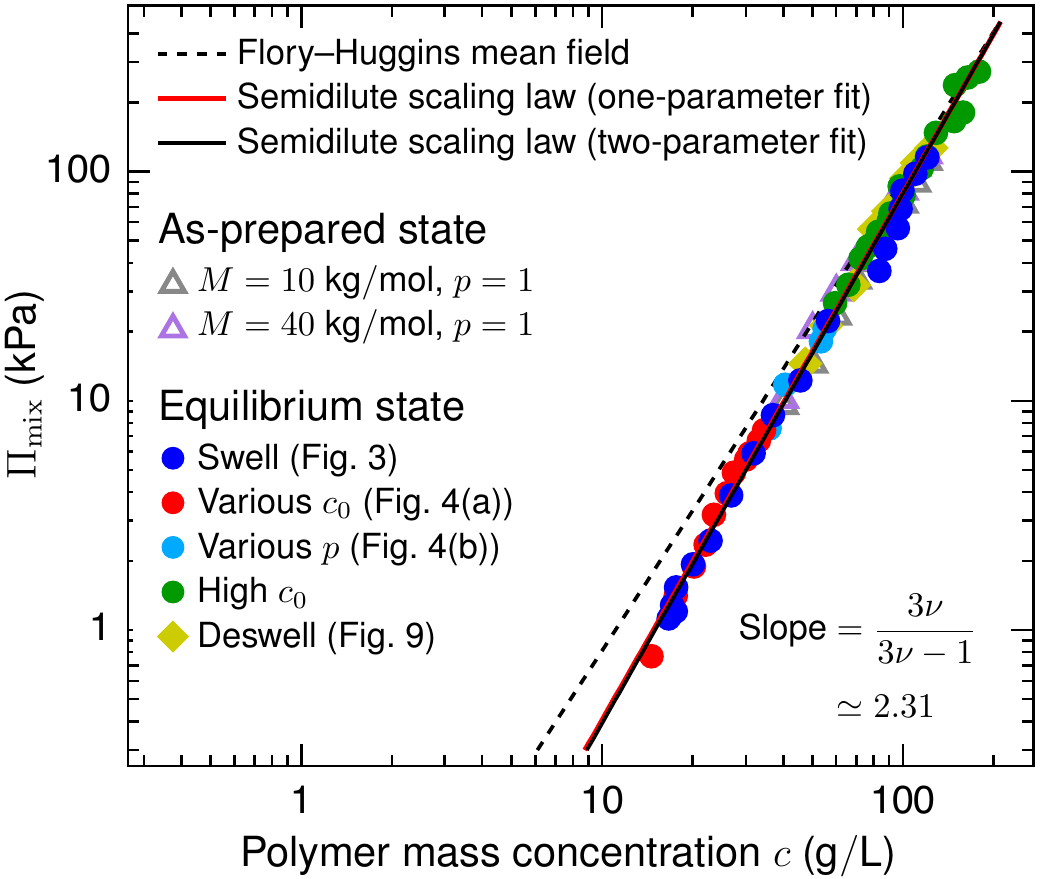}
\caption{
Experimental verification of the semidilute scaling law throughout the quasistatic swelling process.
Log--log plots of the polymer mass concentration ($c$) dependence of $\Pi_\mathrm{mix}$ for the as-prepared state (open gray and purple triangles) and various equilibrium states (filled circles and diamonds) are obtained using Eq.~(\ref{eq:Pos-gel}).
The data of the as-prepared state with molar mass $M=10$ kg$/$mol at $p=1$ are adopted from Ref.~\cite{yasuda2020universal}.
The data of the equilibrium state are the same as those in Figs.~\ref{fig:swelling}, \ref{fig:p,c-dep}, \ref{fig:c*-theorem}, and \ref{fig:deswelling}.
The black dashed curve represents the prediction from the FH theory [Eq.~(\ref{eq:flory-huggins}) with the previously reported value of $\chi=0.426$ \cite{merrill1993partitioning} and $v=18\times10^{-6}\,\mathrm{m^{3}/mol}$].
The black solid line is obtained via a one-parameter ($n^{*}_\mathrm{seg}$) least-square fit to the data of the as-prepared state with the double logarithmic transformation of the semidilute scaling law [Eq.~(\ref{eq:scaling-gel})], i.e., $\ln[\Pi_\mathrm{mix}/(1.1nk_{B}T)]=1.31(\ln n -\ln n^{*}_\mathrm{seg})$.
Here, we used $K=1.1$, $1/(3\nu-1) =1.31$, and $M_{\mathrm{seg}}=44\times10^{-3}$~kg$/$mol as the molar mass of the ethylene glycol unit ($\mathrm{CH_{2}CH_{2}O}$) to calculate $n=cN_{A}/M_\mathrm{seg}$.
The best-fit parameter is $n^{*}_\mathrm{seg}=3.80(8)\times10^{28}\,\mathrm{m}^{-3}$, where the numbers in parentheses indicate the standard deviation in the last digits.
The red solid line is obtained via a two-parameter ($\beta\equiv 1/(3\nu -1)$ and $n^{*}_\mathrm{seg}$) least-square fit to the data in Fig.~\ref{fig:scaling-gel} with the double logarithmic transformation of the semidilute scaling law [Eq.~(\ref{eq:scaling-gel})], i.e., $\ln[\Pi_\mathrm{mix}/(1.1nk_BT)]=\beta(\ln n -\ln n^{*}_\mathrm{seg})$.
The best-fit parameter set is $n^{*}_\mathrm{seg}=3.73(23)\times10^{28}\,\mathrm{m}^{-3}$ and $\beta=1.302(20)$.
}
\label{fig:scaling-gel}
\end{figure}

\section{Semidilute scaling law for $\Pi_\mathrm{mix}$}
\label{sec:Semidilute}

We investigated whether the FH theory or the semidilute scaling law is more accurate in describing $\Pi_\mathrm{mix}$ throughout the quasistatic swelling process.
Figure~\ref{fig:scaling-gel} shows the polymer mass concentration ($c$) dependence of $\Pi_\mathrm{mix}$ for all samples considered in this study, involving the as-prepared, swelling (and deswelling, described in Appendix~\ref{App:deswelling}), and equilibrium swollen states.
In addition, it shows the FH theory [Eq.~(\ref{eq:flory-huggins})] with the previously reported value of $\chi=0.426$ \cite{merrill1993partitioning} and the semidilute scaling law [Eq.~(\ref{eq:scaling-gel})] with the one-parameter least-square fit of $n^{*}_\mathrm{seg}$ to the data of the as-prepared state.
The best-fit parameter, $n^{*}_\mathrm{seg}=3.80(8)\times10^{28}\,\mathrm{m}^{-3}$, is obtained from $\Pi_\mathrm{mix}$ of the polymer gels and is approximately identical to that of the prepolymer solutions in the semidilute regime~\cite{yasuda2020universal} (see Appendix~\ref{App:comparison}).
Here, the number in parentheses is the standard deviation in the last digits.
All the samples (i.e., as-prepared, swelling, deswelling, and equilibrium swollen states with various network conditions of $p$, $c_{0}$, and $M$) conform to the semidilute scaling law with an exponent of $3\nu/(3\nu-1)\simeq2.31$ over a wide concentration range. 
The semidilute scaling law is universal for a large variety of chemically different polymer networks and solvents, as shown in Appendix~\ref{App:previous} using the data available from previous studies \cite{munch1977inelastic,urayama1996elastic,urayama1996crossover,candau1981experimental,horkay1989effects}.
Further, the FH theory fails to describe $\Pi_\mathrm{mix}$ for low concentrations.
This result is consistent with the fact that the FH theory is justified at high concentrations because it is a mean-field theory.
Notably, the difference in $\Pi_\mathrm{mix}$ at low concentrations leads to large deviations in the prediction of the equilibrium volume swelling ratio $Q_\mathrm{eq} \equiv V_\mathrm{eq}/V_{0}$ in the equilibrium swollen state (see Sec.~\ref{sec:prediction}).\\

\begin{table}[t!]
\caption{Universal critical exponent $\nu$ of self-avoiding walks evaluated by different methods.
}
\label{tab:nu}
\begin{ruledtabular}
\begin{tabular}{lll}
References & Method & \, \\
\hline
Ref.~\cite{clisby2016high} & Monte Carlo simulations  & $\nu=0.5875970(4)$ \\
Ref.~\cite{kompaniets2017minimally} &$\epsilon$-expansion at order $\epsilon^6$ &  $\nu=0.5874(3)$ \\
Ref.~\cite{shimada2016conformal} & Conformal bootstrap &  $\nu=0.5877(12)$ \\
\hline
Ref.~\cite{devanand1991asymptotic} & Light scattering of solutions &  $\nu=0.583(3)$ \\
Ref.~\cite{yamamoto1971more} & Light scattering of solutions & $2\nu\simeq 1.17$ \\
Ref.~\cite{fukuda1974solution} & Light scattering of solutions & $2\nu\simeq 1.16$ \\
Ref.~\cite{miyaki1978excluded} & Light scattering of solutions &  $2\nu=1.19(1)$ \\
This study & Swelling experiments of gels &  $\nu\simeq 0.589$\\ 
\end{tabular} 
\end{ruledtabular}
 \end{table}
 
Assuming that the semidilute scaling law [Eq.~(\ref{eq:scaling-gel})] is a fundamental principle governing $\Pi_\mathrm{mix}$ in polymer gels, we can experimentally determine the universal critical exponent $\nu$ of the SAW \cite{de1979scaling,guida1998critical,pelissetto2002critical} from the data of the $c$ dependence of $\Pi_\mathrm{mix}$ in Fig.~\ref{fig:scaling-gel}.
From the two-parameter [$\beta\equiv 1/(3\nu -1)$ and $n^{*}_\mathrm{seg}$] least-square fit to all the data in Fig.~\ref{fig:scaling-gel} with Eq.~(\ref{eq:scaling-gel}), we obtain the best-fit parameter set as $n^{*}_\mathrm{seg}=3.73(23)\times10^{28}\,\mathrm{m}^{-3}$ and $\beta=1.302(20)$.
This result corresponds to $\nu=0.5893(38)$, which is consistent with the previously reported values $\nu=0.5875970(4)$ and $\nu=0.5874(3)$ obtained from Monte Carlo calculations~\cite{clisby2016high} and the $\epsilon$ expansion method~\cite{kompaniets2017minimally}, respectively.
In addition, the obtained value $\nu\simeq 0.589$ is comparable to the previous experimental results of $\nu=0.583(3)$ \cite{devanand1991asymptotic}, $2\nu=1.17$ \cite{yamamoto1971more}, $2\nu=1.16$ \cite{fukuda1974solution}, and $2\nu=1.19(1)$ \cite{miyaki1978excluded}, which were measured by the molar-mass dependence of the gyration radius in dilute polymer solutions with light scattering.
These results are summarized in Table~\ref{tab:nu}.
Notably, our approach of determining $\nu$ using $\Pi_\mathrm{mix}$ of the polymer gels offers advantages over that of polymer solutions \cite{devanand1991asymptotic,yamamoto1971more,fukuda1974solution,miyaki1978excluded} in terms of the polydispersity effect.
Moreover, the overlap threshold is negligible, because polymer gels comprise infinite polymer networks in the semidilute regime.
Therefore, we conclude that measuring the polymer mass concentration dependence of $\Pi_\mathrm{mix}$ of polymer gels provides a new method to experimentally determine the universal critical exponent $\nu$ of the SAW.\\

\begin{figure}[t!]
\centering
\includegraphics[width=\linewidth]{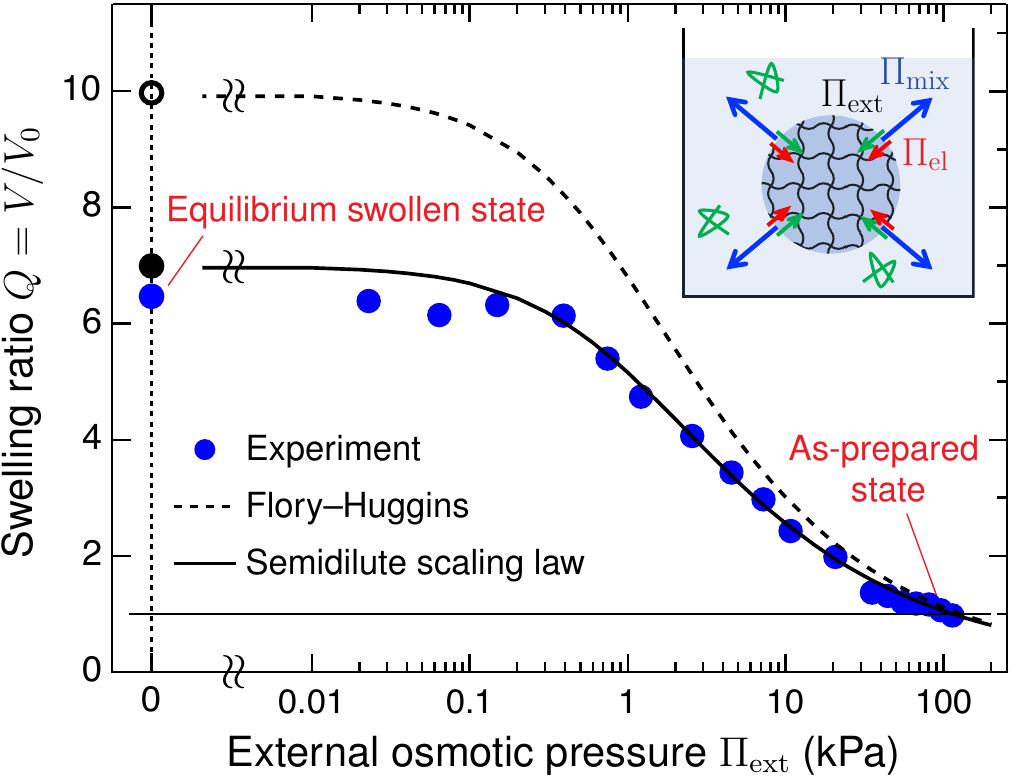}
\caption{
Volume swelling ratio ($Q$) throughout the quasistatic swelling process.
We measured $Q$ by gradually tuning $\Pi_\mathrm{ext}$ (schematic).
The blue circles are the experimental results with $M=40$ kg$/$mol for $c_{0}=120$ g$/$L at $p=0.6$.
The black dashed and solid curves are the predictions from Eq.~(\ref{eq:flory-rehner}) using the FH theory [Eq.~(\ref{eq:flory-huggins}) with $\chi=0.426$ \cite{merrill1993partitioning} and $v=18\times10^{-6}\,\mathrm{m^{3}/mol}$] and the semidilute scaling law [Eq.~(\ref{eq:scaling-gel}) with $n^{*}_\mathrm{seg}=3.80\times10^{28}$~$\mathrm{m}^{-3}$ and $\nu=0.588$], respectively.
In these predictions, we use the polymer mass concentration $c_{0}$ and experimentally measured shear modulus $G_{0}$ at the as-prepared state ($Q=1$).
For the equilibrium swollen state in a pure solvent ($\Pi_\mathrm{ext}=0$),
the experimental result, semidilute scaling law, and FH theory predictions are $Q=6.47, 6.99$, and $9.98$, respectively.
}
\label{fig:swelling}
\end{figure}

\section{Prediction of the quasistatic swelling process}
\label{sec:prediction}

Figure~\ref{fig:swelling} demonstrates that Eq.~(\ref{eq:flory-rehner}) together with the semidilute scaling law [Eq.~(\ref{eq:scaling-gel})], rather than the FH mean-field theory [Eq.~(\ref{eq:flory-huggins})], accurately predicts the experimental value of the volume swelling ratio $Q \equiv V/V_{0}$ throughout the quasistatic swelling process (blue arrow).
The explicit forms of these equations and the typical changes in $\Pi_\mathrm{mix}$ and $\Pi_\mathrm{el}$ are provided in Appendix~\ref{App:equations}.
As shown in Fig.~\ref{fig:scaling-gel}, $\Pi_\mathrm{mix}$ estimated by the FH theory and that by the semidilute scaling law are in agreement at high mass concentrations.
Thus, at high mass concentrations (corresponding to $Q \simeq 1$), the predictions of $Q$ using the FH theory and semidilute scaling law are consistent.
By contrast, $\Pi_\mathrm{mix}$ estimated by the FH theory and that by the semidilute scaling law predict different values in the low mass concentrations, which correspond to low $\Pi_\mathrm{ext}$ in Fig.~\ref{fig:swelling}.
Although the difference in $\Pi_\mathrm{mix}$ in Fig.~\ref{fig:scaling-gel} is small, the difference in $Q$ in Fig.~\ref{fig:swelling} is significantly large.
The deviation of the FH theory from the experimental value of $Q$ in the low $\Pi_\mathrm{ext}$ range may be attributed to the increase in spatial concentration fluctuations at low concentrations, which increases the deviation from the mean-field approximation.\\

The superiority of the semidilute scaling law is also established by the experimental result that $n^*_\mathrm{seg}$ for the polymer gels and prepolymer solutions are approximately identical (see Appendix~\ref{App:comparison}), whereas previously reported values of the FH interaction parameters $\chi$ for polymer gels and polymer solutions exhibit significant inconsistencies \cite{merrill1993partitioning,pedersen2005temperature,venohr1998static,strazielle1968light}.
Thus, by measuring $n^{*}_\mathrm{seg}$ from the polymer solutions, we can determine $\Pi_\mathrm{mix}$ and accurately predict the equilibrium swelling ratio of polymer gels.
Conversely, by measuring $n^{*}_\mathrm{seg}$ from $\Pi_\mathrm{mix}$ of the polymer gels, we can determine the osmotic equation of state of the prepolymer solutions \cite{des1975lagrangian,de1979scaling,noda1981thermodynamic,ohta1982conformation,higo1983osmotic}.
However, applying the FH theory to $\Pi_\mathrm{mix}$ of (dilute) polymer gels does not yield an accurate $\chi$.
These results indicate that the semidilute scaling law is superior to the FH theory in terms of predicting the entire quasistatic swelling process of polymer gels.\\

Figures~\ref{fig:p,c-dep}(a) and (b) further experimentally demonstrate the superiority of the semidilute scaling law over the FH mean-field theory in accurately predicting the experimentally measured volume swelling ratio $Q$ in the equilibrium swollen state with the various network conditions of the connectivity $p$ and polymer mass concentration in the as-prepared state $c_{0}$.
The semidilute scaling law accurately predicted $Q$ under all conditions within the experimental error, whereas the FH theory with the constant $\chi$ predicted a significantly higher $Q$.
These results indicate the superiority of the semidilute scaling law to the FH theory in predicting polymer gel swelling over a wide concentration range at any equilibrium condition.
The FH theory is generally inadequate for application in polymer gels; instead, the semidilute scaling law governs polymer gels.

\begin{figure}[t!]
\centering
\includegraphics[width=\linewidth]{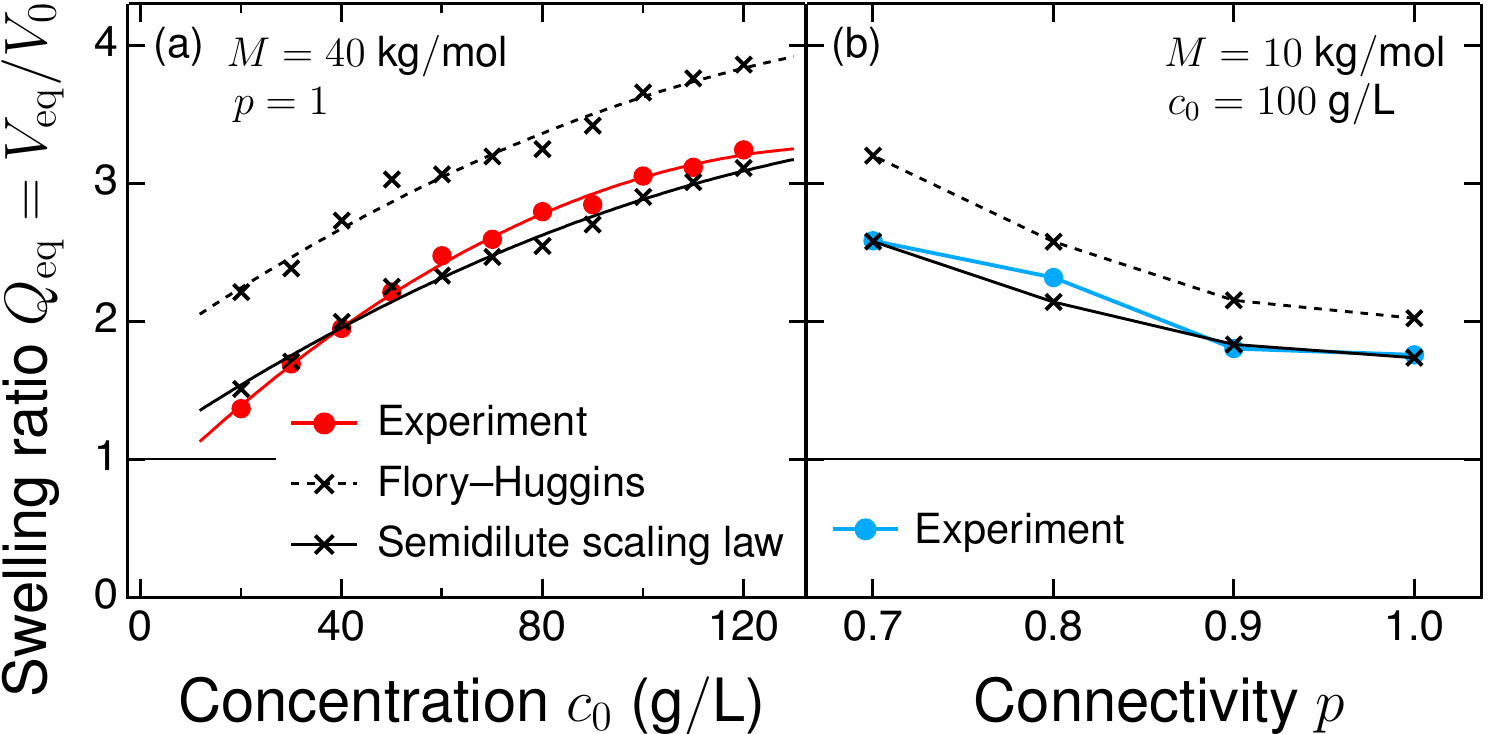}
\caption{
Equilibrium swelling ratio at the equilibrium swollen state with various network conditions in a pure solvent ($\Pi_\mathrm{ext}=0$).
The red and light blue circles denote the experimental results 
(a) with $M=40$ kg$/$mol for $c_{0}=20$ -- $120$ g$/$L at $p=1$ and (b) with $M=10$ kg$/$mol for $c_{0}=100$ g$/$L at $p=0.7, 0.8, 0.9,$ and $1$, respectively.
The black solid and dashed curves are the predictions from Eq.~(\ref{eq:flory-rehner}) using the FH theory [Eq.~(\ref{eq:flory-huggins})] and the semidilute scaling law [Eq.~(\ref{eq:scaling-gel})], respectively.
In these predictions, we use the same values of $v$, $\chi$, $\nu$, and $n^{*}_\mathrm{seg}$ as those in Fig.~\ref{fig:swelling}.
Moreover, we use the polymer mass concentration ($c_{0}$) and experimentally measured shear modulus ($G_{0}$) for each sample at the as-prepared state ($Q=1$).
}
\label{fig:p,c-dep}
\end{figure}

\begin{figure}[t!]
\centering
\includegraphics[width=\linewidth]{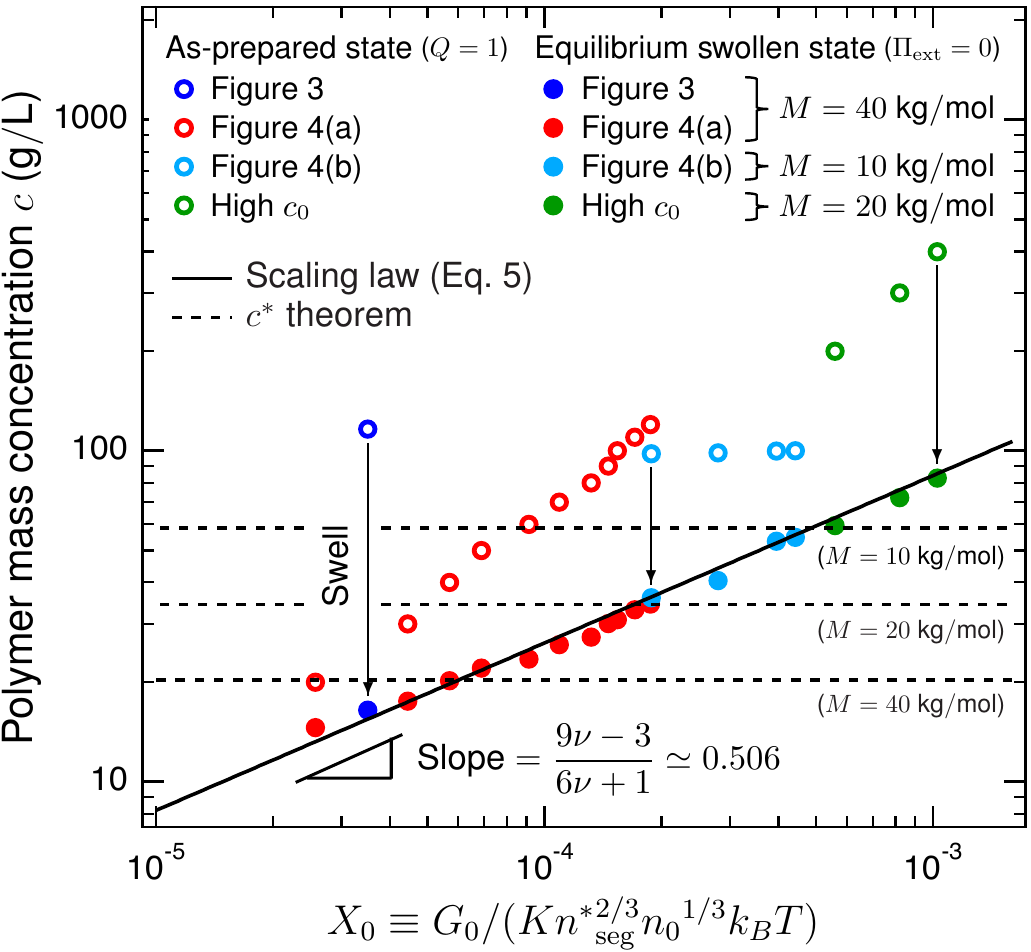}
\caption{
Semidilute scaling law (black solid line) for determining the equilibrium concentrations [Eq.~(\ref{eq:equilibrium-c})].
The open circles are the polymer mass concentrations of polymer gels at the as-prepared state ($Q=1$).
The dimensionless parameter $X_{0}\equiv G_{0}/(K {n^{*}}^{2/3}_\mathrm{seg} n_{0}^{1/3}k_{B}T)$ at the as-prepared state determines the equilibrium concentrations $c_\mathrm{eq}$ (filled circles) of polymer gels in a pure solvent ($\Pi_\mathrm{ext}=0$) with $n^{*}_\mathrm{seg}=3.80\times10^{28}$~$\mathrm{m}^{-3}$.
The experimental data are the same as those in Figs.~\ref{fig:swelling} and \ref{fig:p,c-dep}.
De Gennes' $c^{*}$ theorem predicts that $c_\mathrm{eq}$ is constant (gray dashed lines) for each $M$; this fails to explain the dependence of $c_\mathrm{eq}$ on the as-prepared states ($c_0$ and $G_0$). 
To plot the gray dashed lines, we use $c_\mathrm{eq}=58.5$, $34.4$, and $20.3$~g/L for $M=10$, $20$, and $40$~kg/mol, respectively.
Here, $58.5$~g/L is the overlap concentration $c^{*}$ of prepolymer solutions with $M=10$~kg/mol \cite{yasuda2020universal}.
Moreover, $34.4$ and $20.3$~g/L is calculated from $c^{*}\propto M^{1-3\nu}$ in the polymer solutions.
}
\label{fig:c*-theorem}
\end{figure}

\section{Scaling law for equilibrium concentration beyond de Gennes' $c^{*}$ theorem}
\label{sec:c* theorem}

Based on the experimental results, we propose a scaling law that predicts $c_\mathrm{eq}$ (i.e., the polymer mass concentration at $\Pi_\mathrm{tot}=0$), while contradicting the predictions made by the $c^*$ theorem.
The semidilute scaling law [Eq.~(\ref{eq:scaling-gel})] together with Eq.~(\ref{eq:flory-rehner}) and $\Pi_\mathrm{tot}=0$ yields 
\begin{equation}
n_\mathrm{eq}=n^{*}_\mathrm{seg} X_0^{(9\nu-3)/(6\nu+1)},
\label{eq:equilibrium-c}
\end{equation}
where $X_{0}\equiv G_{0}/(K {n^{*}}^{2/3}_\mathrm{seg} n_{0}^{1/3}k_{B}T)$ is a dimensionless parameter determined by the parameters of the as-prepared state ($G_0$ and $n_0$).
Figure~\ref{fig:c*-theorem} shows that the equilibrium concentrations ($c_\mathrm{eq}$) in Figs.~\ref{fig:swelling} and \ref{fig:p,c-dep} are governed by 
Eq.~(\ref{eq:equilibrium-c}) rather than by the $c^{*}$ theorem.
Each gel swells toward the equilibrium swollen state from the as-prepared state at a constant $X_{0}$ (black arrows).
All gels swollen in a pure solvent (filled symbols) with various network conditions of $p$, $c_{0}$, and $M$ exhibit the scaling law of Eq.~(\ref{eq:equilibrium-c}) with an exponent of $0.506$, whereas the $c^{*}$ theorem fails to predict the $G_{0}$ and $c_{0}$ dependences of $c_\mathrm{eq}$.
Thus, we can determine the equilibrium concentration and volume of the swollen polymer gels by measuring $n^{*}_\mathrm{seg}$ from Eq.~(\ref{eq:scaling-gel}) and calculating $X_{0}$.\\

The present results also clarify the inaccuracy of the interpretation of the experimental results that had affirmed the $c^{*}$ theorem.
The scaling relation of Young's modulus $E = 3G \sim \phi^{3\nu/(3\nu-1)}$ \cite{munch1977inelastic,urayama1996elastic,urayama1996crossover,candau1981experimental,horkay1989effects} in the equilibrium swollen state serves as the experimental evidence of the $c^{*}$ theorem.
However, this observation is only a necessary condition for the $c^{*}$ theorem, not a sufficient condition.
This observation is a consequence of the semidilute scaling law [Eq.~(\ref{eq:scaling-gel})], because $E \sim |\Pi_\mathrm{el}| = \Pi_\mathrm{mix} \sim c^{3\nu/(3\nu-1)}$ in the equilibrium swollen state ($\Pi_\mathrm{tot}=0$), as shown in Fig.~\ref{fig:scaling-gel}.
A major deficiency of the $c^{*}$ theorem is that the network elastic contribution $\Pi_\mathrm{el}$, which depends on the as-prepared state, is neglected in determining the equilibrium concentrations of the polymer gels, assuming complete disinterpenetration of the subchains in a swollen network.
By contrast, Eq.~(\ref{eq:equilibrium-c}) includes the elastic contribution, as follows:
$\Pi_\mathrm{el} = -G_{0}\left(c/c_{0}\right)^{1/3}$.
This can be used to predict the equilibrium concentrations of polymer gels in a pure solvent as the scaling law of the one-parameter $X_{0}$.
Our results reveal the scaling law governing $c_\mathrm{eq}$ by (i) assuming the semidilute scaling law, rather than the FH mean-field theory, governs $\Pi_\mathrm{mix}$; 
(ii) changing the definition of the polymer mass concentration to extend the universality of the semidilute scaling law to higher concentrations;
and (iii) using experimentally measured $G_{0}$ in the as-prepared state for determining the network elastic contributions $\Pi_\mathrm{el}$ of the swollen polymer gels, considering the negative energetic elasticity \cite{yoshikawa2021negative,Sakumichi2021,fujiyabu2021temperature,Shirai2022}.

\section{Concluding remarks}
\label{sec:Conclusion}

We experimentally measured $\Pi_\mathrm{mix}$ of chemically crosslinked polymer gels with precisely controlled homogeneous network structures throughout the quasistatic swelling process.
We find that $\Pi_\mathrm{mix}$ is governed by the semidilute scaling law of polymer solutions [Eq.~(\ref{eq:scaling-gel})] and enables accurate predictions of the swelling ratio of polymer gels throughout the quasistatic swelling process.
By contrast, the conventional FH mean-field theory [Eq.~(\ref{eq:flory-huggins})] fails to describe $\Pi_\mathrm{mix}$ for low concentrations and predicts a significantly higher equilibrium swelling ratio.
Assuming the semidilute scaling law as a fundamental principle (semidilute principle) for polymer gels, we obtain $\nu\simeq 0.589$, which indicates that measuring the polymer mass concentration dependence of $\Pi_\mathrm{mix}$ provides a new method for experimentally determining the universal critical exponent $\nu$ of the SAW.
Furthermore, the semidilute principle, together with the experimentally measured shear modulus in the as-prepared state $G_{0}$, yields the universal scaling law [Eq.~(\ref{eq:equilibrium-c})] that determines the equilibrium concentrations of polymer gels in a pure solvent, thereby demonstrating that de Gennes' $c^{*}$ theorem cannot explain the experimental results.\\

The scaling law [Eq.~(\ref{eq:equilibrium-c})] is considered a general law that can be applied not only to the present gel but also to other polymer gels that are electrically neutral binary mixtures in good solvents. 
Moreover, the semidilute principle, which has been established for electrically neutral gels, is expected to be a firm basis for the prediction of swelling of charged polymer gels such as living organisms.
Because Eq.~(\ref{eq:flory-rehner}) is a sum of $\Pi_\mathrm{mix}$ and $\Pi_\mathrm{el}$ without any cross-correlations \cite{horkay2000osmotic,ricka1984swelling,tang2020swelling,jia2021theory}, the semidilute principle for electrically neutral gels can be extended to charged gels by simply adding an excess pressure term to account for electric charge. 
Therefore, the semidilute principle facilitate fundamental understanding of polymer gels, and will find boarder applications in fields such as life science \cite{peppas1997hydrogels,xinming2008polymeric}, biomaterials \cite{porchet1998clinical,hayashi2017fast}, petroleum extraction \cite{karacan2007swelling,makitra2011determination}, and cooking \cite{nelson2020mathematical}.

\begin{acknowledgments}
We would like to thank Nobu C. Shirai for the useful comments.
This work was supported by the Japan Society for the Promotion of Science (JSPS) through the Grants-in-Aid for 
Early Career Scientists Grant No.~19K14672 to N.S.,
Scientific Research (B) Grant No.~22H01187 to N.S.,
JSPS Research Fellows Grant No.~202214177 to T.Y.,
and Scientific Research (A) Grant No.~21H04688 to T.S.,
and Transformative Research Area Grant No.~20H05733 to T.S.
This work was also supported by the Japan Science and Technology Agency (JST) CREST Grant No.~JPMJCR1992 to T.S.
\end{acknowledgments}

\section*{Author contributions}

N.S., T.Y., and T.S. conceived, designed, and discussed the research.
N.S. developed the theoretical framework.
T.Y. designed and performed the experiments.
N.S. and T.Y. analyzed and interpreted the results and wrote the manuscript.
T.S. supervised the whole research.

%\section*{Competing interests:}
%The authors declare no competing interests.

%\section*{Data and materials availability:}
%All data used in this study are available from the corresponding authors upon reasonable request.

%\section*{Additional information:}
%Correspondence and requests for materials should be addressed to N.S. or T.S.

\appendix

\section{Measured shear modulus at the as-prepared state}
\label{App:shear modulus}

\begin{figure}[b!]
\centering
\includegraphics[width=\linewidth]{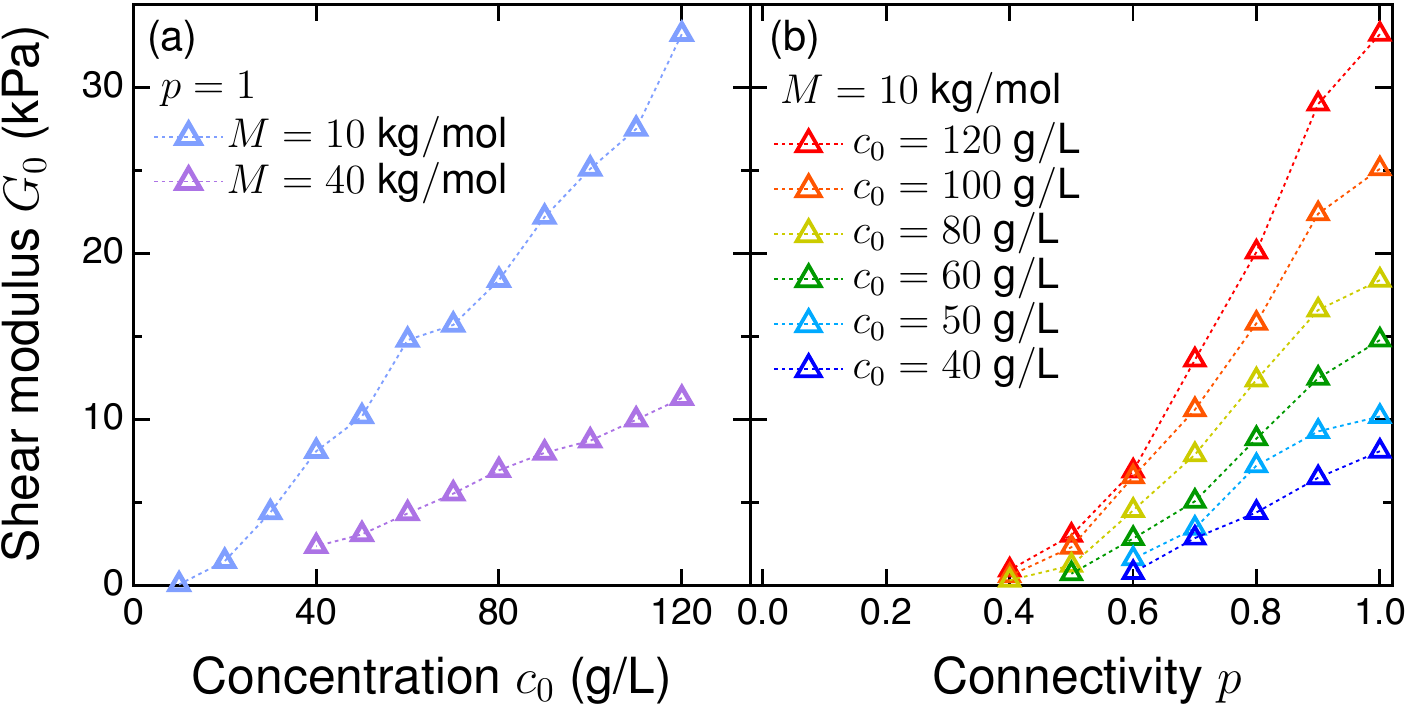}
\caption{
Dependence of shear modulus at the as-prepared state on polymer mass concentration and connectivity.
(a) Dependence of $G_{0}$ on the polymer concentration $c_{0}=10$ -- $120$ g$/$L with the molar mass $M=10$ and $40$ kg$/$mol at $p=1$.
(b) Dependence of $G_{0}$ on the connectivity $p=2s=0.4, 0.5, 0.6, 0.7, 0.8, 0.9$, and $1$ for the various polymer concentration $c_{0}=40$ -- $120$ g$/$L with molar mass $M=10$ kg$/$mol.
}
\label{fig:rheometer}
\end{figure}

Figures~\ref{fig:rheometer}(a) and (b) show the dependence of $G_{0}$ on the polymer concentration ($c$) and connectivity ($p$), respectively.
In addition, to obtain the results shown in Fig.~\ref{fig:scaling-gel} (samples corresponding to the green and blue filled circles), we measured $G_{0}=40.0$, $67.2$, and $92.3$ kPa for $c_{0}=200$, $300$, and $400$ g$/$L, respectively, with $M=20$ kg$/$mol at $p=1$, and $G_{0}=2.1$ kPa for $c_{0}=120$ g$/$L with $M=40$ kg$/$mol at $p=0.6$, which is not shown in Fig.~\ref{fig:rheometer}.
For a low $c_{0}$ or low $p$, swelling experiments cannot be performed, because gels lose their shape and weight owing to their low elasticity.

\begin{figure}[b!]
\centering
\includegraphics[width=\linewidth]{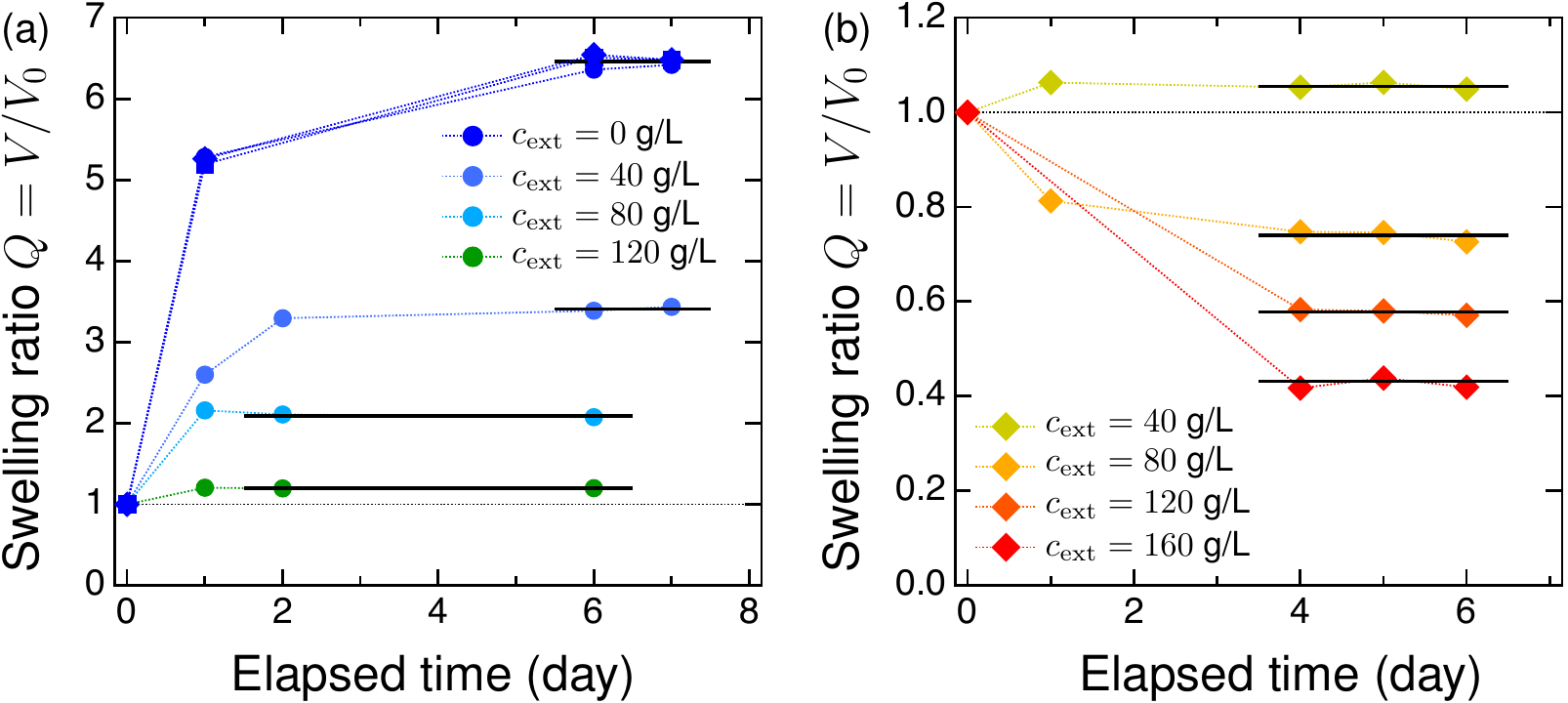}
\caption{
Time courses of swelling ratio in the (a) swelling and (b) deswelling experiments.
The sample conditions for (a) and (b) are the same as those for Fig.~\ref{fig:swelling} (blue circles) and Fig.~\ref{fig:deswelling} (yellow diamonds), respectively.
Horizontal black lines indicate equilibrium swelling ratio.
(a) We set $c_\mathrm{ext}=0, 40, 80,$ and $120$ g$/$L for the gel samples with $M=40$ kg$/$mol for $c_{0}=120$ g$/$L at $p=0.6$.
For $c_\mathrm{ext}=0$ (blue symbols), the three representative experiments agreed with each other.
(b) We set $c_\mathrm{ext}=40, 80, 120,$ and $160$ g$/$L for the gel samples with $M=10$ kg$/$mol for $c_{0}=50$ g$/$L at $p=1$.
}
\label{fig:relaxation}
\end{figure}

\section{Equilibrium state in swelling experiments and its verification}
\label{App:achievement}

We evaluated the polymer mass concentration $c$ at the equilibrium states in the swelling experiments as
\begin{equation}
c=\frac{c_{0}}
{Q+(Q-1)c_{0}/\rho_\mathrm{PEG}},
\label{eq:concentration}
\end{equation}
where the volume swelling ratio $Q \equiv V/V_{0}$ is calculated as
\begin{equation}
Q=\frac{W}{W_0}+
\left(
\frac{W}{W_0}-1
\right)
\frac{(\rho_\mathrm{PEG}/\rho_\mathrm{water}-1)c_{0}}{\rho_\mathrm{PEG}+c_{0}},
\label{eq:swelling-ratio}
\end{equation}
using experimentally measured $W_{0}$ and $W$.
Here, $\rho_\mathrm{PEG}\simeq 1129\,\mathrm{kg}/\mathrm{m}^{3}$ and $\rho_\mathrm{water}\simeq 1000\,\mathrm{kg}/\mathrm{m}^{3}$ are the polymer (PEG) and solvent densities, respectively.
Further, we estimated the amount of polymeric clusters slightly leaked from the gel samples with $p<1$, to evaluate $c$ of the gels immersed in the solvents with a high accuracy (see Appendix~\ref{App:outflow}).
Notably, we define the polymer mass concentrations ($c_{0}$, $c$, and $c_\mathrm{eq}$) as the prepolymer weight divided by the solvent volume, rather than the solution volume, to extend the universality of the semidilute scaling law [Eq.~(\ref{eq:scaling-gel})] to higher concentrations.
(Further details are provided in Sec.~S1 in the Supplemental Material in Ref.~\cite{yasuda2020universal}.)
Similarly, to calculate $\Pi_\mathrm{mix}$ in the FH mean-field theory given by Eq.~(\ref{eq:flory-huggins}), we evaluate the polymer volume fraction $\phi$ as
\begin{equation}
\phi=\frac{c_{0}/\rho_\mathrm{PEG}}
{Q(1+c_{0}/\rho_\mathrm{PEG})}.
\label{eq:phi}
\end{equation}

We ensure that each gel sample reached equilibrium in the swelling experiments, as evidenced by the swelling and deswelling processes of the polymer gels to the equilibrium state illustrated in Figs.~\ref{fig:relaxation}(a) and (b), respectively.
To reach equilibrium (horizontal black lines), approximately one week was required.
Here, we determined that the equilibrium was reached when the volume swelling ratio $Q$ remained constant for two or three days.
We performed three swelling experiments under the same conditions [blue symbols in Fig.~\ref{fig:relaxation}(a) for $c_\mathrm{ext}=0$] to confirm the reproducibility of the time course of $Q$.

\section{Leakage of polymeric clusters from gels fabricated at a low connectivity}
\label{App:outflow}

We evaluate the amount of leakage of polymeric clusters from each gel sample fabricated at a connectivity $p$ such that $p_\mathrm{gel}<p<1$ (stoichiometrically imbalanced mixing).
Here, $p_\mathrm{gel}$ is the connectivity at gel point.
Upon immersing a gel sample (fabricated at $p$) in solvents, the unreacted terminal functional groups yield polymeric clusters that are unconnected to the gel network.
Thus, the polymer mass of the gel samples decreases from pre-immersion ($\mathcal{M}_0$) to post-immersion ($\mathcal{M}_0^{\mathrm{im}}$).
The polymer mass concentration at pre-immersion is $c_{0}\equiv \mathcal{M}_0/V_0$, where $V_0$ is the volume of the gel sample.
Assuming that the semidilute scaling law [Eq.~(\ref{eq:scaling-gel})] holds, we can estimate the polymer mass concentration at post-immersion $c_{0}^{\mathrm{im}}\equiv \mathcal{M}_0^{\mathrm{im}}/V_0$ as
\begin{equation}
c_{0}^{\mathrm{im}} \left(c_{0},p\right)
=c_{0}\cdot
\left[
\frac{\Pi_\mathrm{mix}(p)}{\Pi_{\mathrm{mix}}(1)}
\right]
^{\frac{3\nu-1}{3\nu}},
\label{eq:outflow}
\end{equation}
where ${\Pi_{\mathrm{mix}}{(p)}}$ is the mixing contribution of the osmotic pressure of the gel sample ($p_\mathrm{gel}<p\leq 1$) and ${\Pi_{\mathrm{mix}}{(1)}}$ is that of the as-prepared gel sample fabricated at $p=1$.\\

\begin{figure}[t!]
\centering
\includegraphics[width=\linewidth]{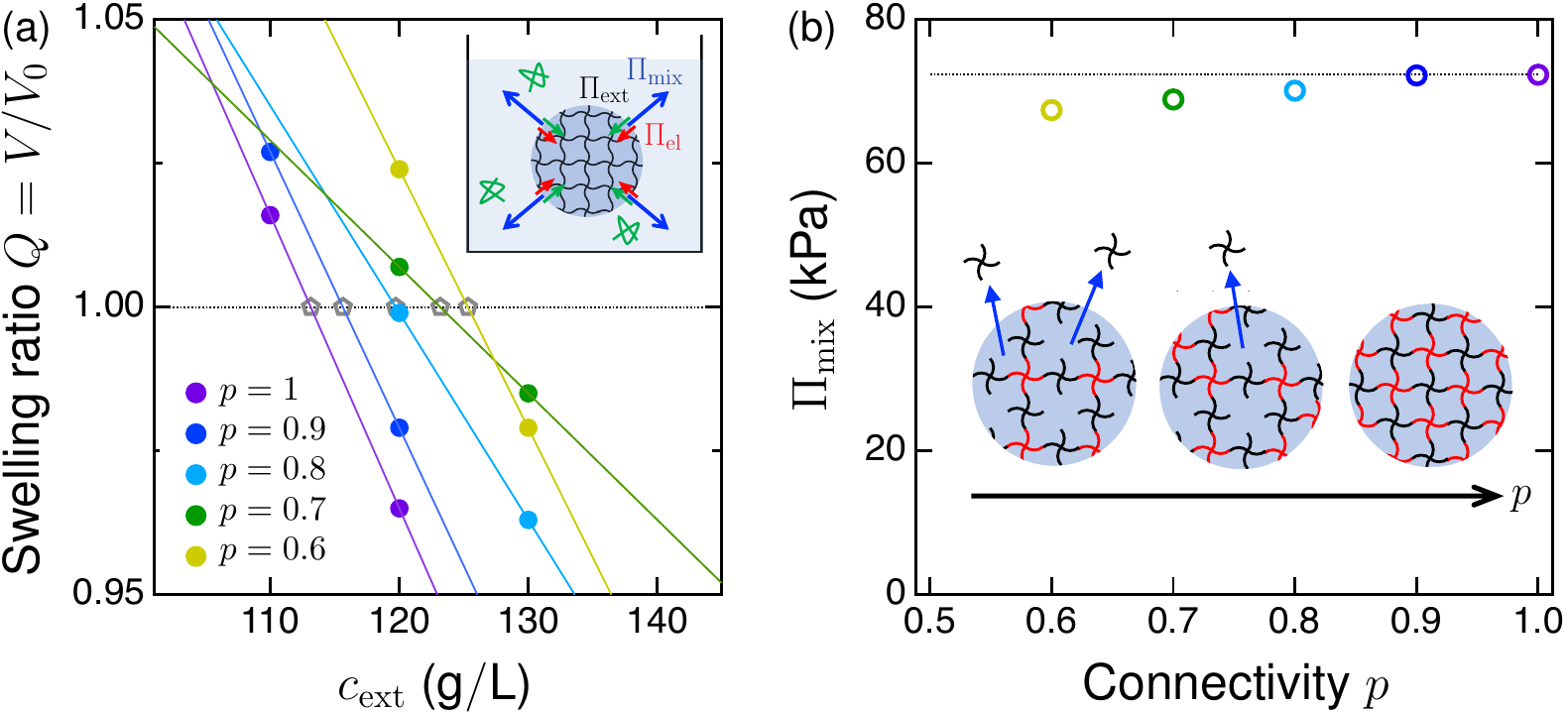}
\caption{
Estimated decrease in polymer mass concentration of polymer gels in the as-prepared state due to the leakage of polymeric (PEG) clusters.
(a) Volume swelling ratio $Q$ and (b) $\Pi_\mathrm{mix}$ of gel samples for the connectivity $p=2s=0.6, 0.7, 0.8, 0.9,$ and $1$ at the equilibrium state in the external polymer (PVP) solutions.
The molar mass and mass concentration of the prepolymer solutions were $M=10$ kg$/$mol and $c_{0}=100$ g$/$L, respectively.
(a) By interpolating $c_\mathrm{ext}$ that yields $Q=1$ for each gel sample (gray pentagons), we can determine $\Pi_\mathrm{mix}$ for each as-prepared gel sample, as $\Pi_\mathrm{mix}=\Pi_\mathrm{ext}+G_{0}$.
(b) We evaluated each $\Pi_\mathrm{mix}$ using the data in (a).
As $p$ decreases, $\Pi_\mathrm{mix}$ also decreases owing to the increase in the leakage of the polymeric clusters.
}
\label{fig:outflow}
\end{figure}

\begin{table}[h!]
\caption{The decrease of polymer mass concentration of the gel samples from pre-immersion ($c_0$) to post-immersion ($c_0^\mathrm{im}$) evaluated by $\Pi_\mathrm{mix}$ for each $p$.
The gel samples are in the as-prepared state with $M=10$~kg$/$mol for $c_{0}=100$~g$/$L and with $M=40$~kg$/$mol for $c_{0}=120$~g$/$L.
}
\label{tab:cim}
\begin{ruledtabular}
\begin{tabular}{cccccccc}
$c_0$ (g$/$L) & \multicolumn{5}{c}{100} & \multicolumn{2}{c}{120} \\ \cline{2-6} \cline{7-8}
$p$ & $0.6$ & $0.7$ & $0.8$ & $0.9$ & $1$ & $0.6$ & $1$\\
\hline
$\Pi_\mathrm{mix}$ (kPa) & $67.4$ & $68.9$ & $70.1$ & $72.2$ & $72.3$ & $110.5$ & $116.1$ \\
$c_0^\mathrm{im}$ (g$/$L) & 97.0 & 97.9 & 98.7 & 99.9 & 100 & 117.4 & 120 \\
\end{tabular} 
\end{ruledtabular}
 \end{table}

By using Eq.~(\ref{eq:outflow}) and measuring the connectivity ($p$) dependence of $\Pi_\mathrm{mix}$ of the gel samples in the as-prepared state, we estimated the amount of leakage of polymeric clusters.
Figure~\ref{fig:outflow}(a) shows the interpolation of $c_\mathrm{ext}$, which yields $Q=1$ for each gel sample with $M=10$ kg$/$mol for $c_{0}=100$ g$/$L at $p=0.6$, $0.7$, $\dots$, and $1.0$ (gray pentagons).
Thus, we determined $\Pi_\mathrm{mix}$ of each as-prepared gel sample as $\Pi_\mathrm{mix}=\Pi_\mathrm{ext}+G_{0}$.
Figure~\ref{fig:outflow}(b) shows the connectivity ($p$) dependence of $\Pi_\mathrm{mix}$ evaluated using the data in Fig.~\ref{fig:outflow}(a).
As $p$ decreases, $\Pi_\mathrm{mix}$ also decreases because of the increase in the amount of leakage of the polymeric clusters (schematic in Fig.~\ref{fig:outflow}(b)).\\

\begin{figure}[b!]
\centering
\includegraphics[width=\linewidth]{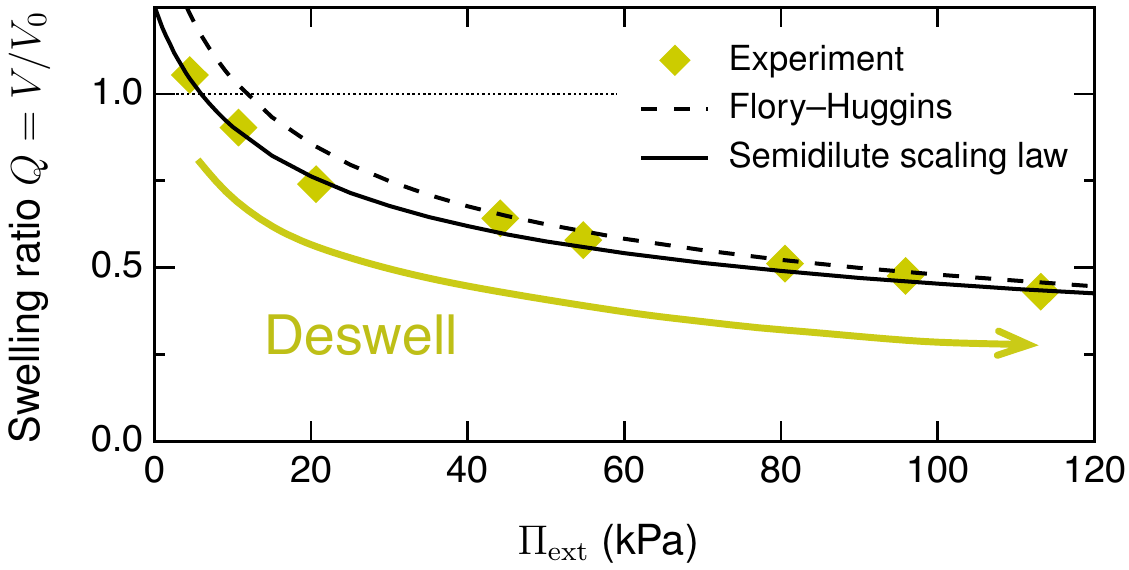}
\caption{
Swelling ratio throughout the quasistatic deswelling process.
We measured $Q$, by gradually tuning $\Pi_\mathrm{ext}$.
The yellow diamonds are the experimental results with $M=10$ kg$/$mol for $c_{0}=50$ g$/$L at $p=1$.
The black solid and dashed curves are the predictions using the FH theory [Eq.~(\ref{eq:Q-FH}) with $\chi=0.426$ \cite{merrill1993partitioning} and $v=18\times10^{-6}\,\mathrm{m^{3}/mol}$] and the semidilute scaling law [Eq.~(\ref{eq:Q-scaling}) with $n^{*}_\mathrm{seg}=3.80\times10^{28}$~$\mathrm{m}^{-3}$ and $\nu=0.588$], respectively.
In these predictions, we use the polymer concentration $c_{0}$ and experimentally measured shear modulus $G_{0}$ at the as-prepared state ($Q=1$).
}
\label{fig:deswelling}
\end{figure}

Table~\ref{tab:cim} shows $c_0^\mathrm{im}$ of the gel samples in the as-prepared state with $M=10$~kg$/$mol for $c_{0}=100$~g$/$L and with $M=40$~kg$/$mol for $c_{0}=120$~g$/$L.
The decreasing rate is $1-c_{0}^{\mathrm{im}}/c_{0}\simeq 0.025$--$0.030$ for these samples at $p=0.6$.
The obtained $c_{0}^{\mathrm{im}}$ was used in Fig.~\ref{fig:scaling-gel} to evaluate the quasistatic swelling process (blue circles) and various connectivities (light blue circles).
We also used the evaluated $c_{0}^{\mathrm{im}}$ to predict $Q$ and $c_\mathrm{eq}$ in Figs~\ref{fig:swelling}, \ref{fig:p,c-dep}, and \ref{fig:c*-theorem}.

\begin{figure}[t!]
\centering
\includegraphics[width=\linewidth]{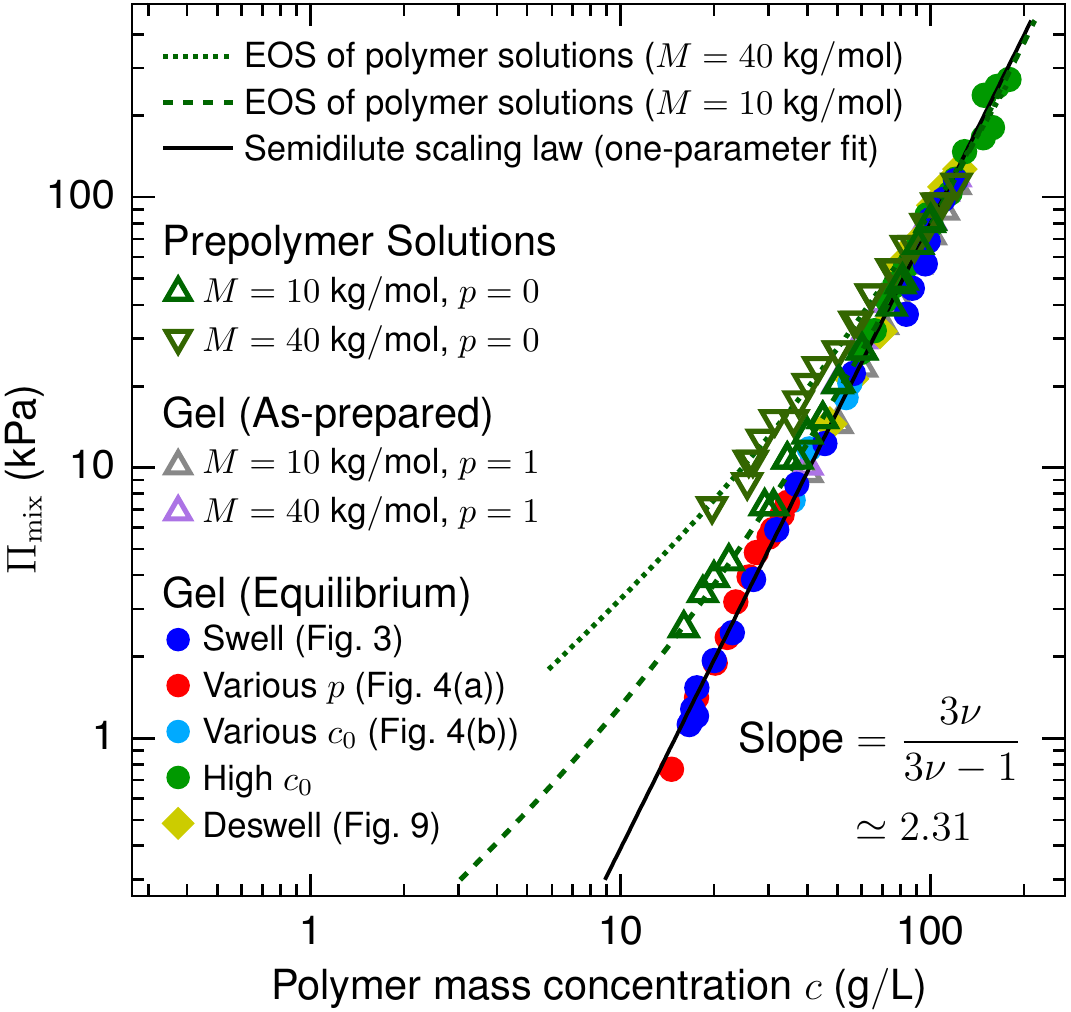}
\caption{
Polymer mass concentration ($c$) dependence of $\Pi_\mathrm{mix}$ of polymer gels, compared with that in the unreacted prepolymer solutions with the molar mass $M=10$ and $40$ kg$/$mol (green triangles). 
The polymer gels (data from Fig.~\ref{fig:scaling-gel}) obeys the semidilute scaling law [Eq.~\ref{eq:scaling-gel}] with the exponent $3\nu/(3\nu-1)\simeq2.31$ for $n^{*}_\mathrm{seg}=3.80\times 10^{28}\,\mathrm{m}^{-3}$ (black solid line).
The prepolymer solutions obeys the universal osmotic equation of state of polymer solutions (green dashed curves) \cite{des1975lagrangian,de1979scaling,noda1981thermodynamic,ohta1982conformation,higo1983osmotic}.
The data of the prepolymer (tetra-PEG MA) solutions are obtained from Ref.~\cite{yasuda2020universal}.}
\label{fig:comparison}
\end{figure}

\begin{figure}[t!]
\centering
\includegraphics[width=\linewidth]{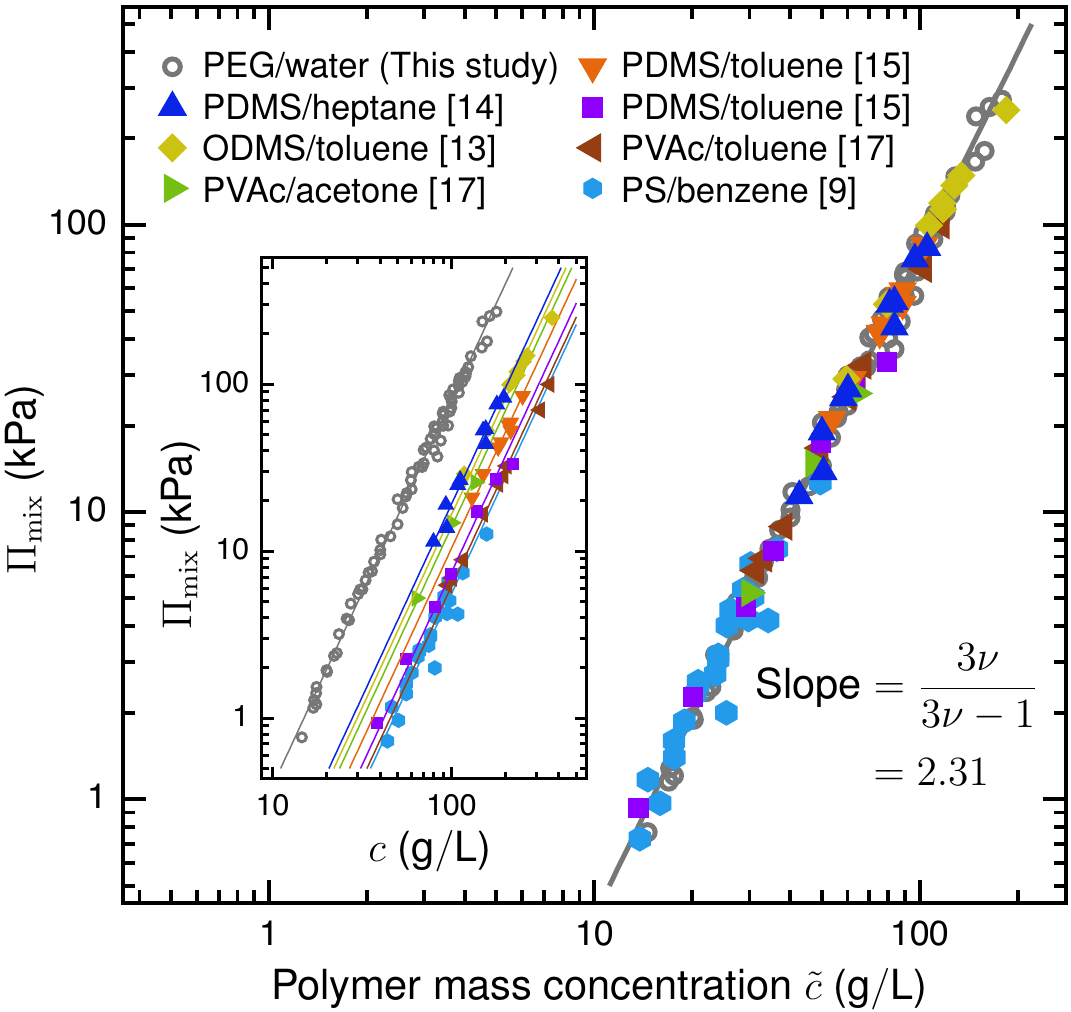}
\caption{
Dependence of $\Pi_\mathrm{mix}$ in this study (open symbols) and previous studies (filled symbols) \cite{munch1977inelastic,urayama1996elastic,urayama1996crossover,candau1981experimental,horkay1989effects} on the polymer mass concentration with the horizontal shift  ($c$ to $\tilde{c}$).
Inset: Polymer mass concentration ($c$) dependence of $\Pi_\mathrm{mix}$.
Each solid line represents a one-parameter least-square fit to the data with the double logarithmic transformation of the semidilute scaling law [Eq.~(\ref{eq:scaling-gel})], i.e., $\ln[\Pi_\mathrm{mix} /(1.1nk_B T)] = 1.31(\ln n - \ln n^*_\mathrm{seg}$).
}
\label{fig:previous}
\end{figure}

\section{Deswelling experiments}
\label{App:deswelling}

In Fig~\ref{fig:deswelling}, we demonstrate that the semidilute scaling law, rather than FH mean-field theory, accurately predicts the volume swelling ratio ($Q\equiv V/V_{0}$) in the quasistatic deswelling process.
Although polymer gels generally swell because $\Pi_\mathrm{tot}=\Pi_\mathrm{mix}+\Pi_\mathrm{el}>0$ in the as-prepared state, we can quasistatically deswell the polymer gels by gradually increasing $\Pi_\mathrm{ext}$ from the as-prepared state via osmotic deswelling \cite{boyer1945deswelling,bastide1981osmotic}.
For the quasistatic deswelling process (yellow arrow), the semidilute scaling law accurately predicts the experimental result ($Q$) under the present experimental conditions ($Q>0.4$ and $c<120$ g$/$L), whereas the FH mean-field theory fails to predict $Q$ around $Q=1$.
This result supports the superiority of the semidilute scaling law over the FH mean-field theory for $\Pi_\mathrm{mix}$ of polymer gels.
However, we note that there is a limitation of the semidilute scaling law for $\Pi_\mathrm{mix}$ in the quasistatic deswelling process because polymer gels will be beyond the semidilute regime at a significantly lower $Q$.

\section{Comparison of osmotic pressure of polymer gels and prepolymer solutions}
\label{App:comparison}

To evaluate the consistency of $n^{*}_\mathrm{seg}$ in Eq.~(\ref{eq:scaling-gel}), we experimentally compared the polymer concentration ($c$) dependence of $\Pi_\mathrm{mix}$ in Fig.~\ref{fig:scaling-gel} with that of the unreacted prepolymer solutions ($p=0$) \cite{yasuda2020universal}. 
Figure~\ref{fig:comparison} demonstrates that $\Pi_\mathrm{mix}$ of the polymer gels exhibits the semidilute scaling law [Eq.~(\ref{eq:scaling-gel})] with an exponent of $3\nu/(3\nu-1)\simeq2.31$.
By contrast, the prepolymer solutions obey the universal osmotic equation of state (green dashed curves) \cite{des1975lagrangian,de1979scaling,noda1981thermodynamic,ohta1982conformation,higo1983osmotic}, because prepolymers have finite molar mass and in the dilute regime ($c < c^{*}$) for a low $c$.
Notably, for a high $c$, $\Pi_\mathrm{mix}$ of prepolymer solutions (green dashed curves) is asymptotic to the semidilute scaling law of polymer gels (black solid line).
The $\Pi_\mathrm{mix}$ of polymer gels is approximately identical to that of prepolymer solutions in the semidilute regime ($c \gg c^{*}$).
Thus, the semidilute scaling law [Eq.~(\ref{eq:scaling-gel})] for polymer gels is nothing but that for prepolymer solutions.
Based on this result, we can determine $n^{*}_\mathrm{seg}$ in Eq.~(\ref{eq:scaling-gel}) by measuring the osmotic equation of state of the prepolymer solutions.

\newpage

\section{Semidilute scaling law for $\Pi_\mathrm{mix}$ in previous studies}
\label{App:previous}

Figure~\ref{fig:previous} shows the polymer mass concentration dependence of $\Pi_\mathrm{mix}$ in this study (gray open circles) and previous studies (filled symbols) \cite{munch1977inelastic,urayama1996elastic,urayama1996crossover,candau1981experimental,horkay1989effects} and demonstrates that the semidilute scaling law [Eq.~(\ref{eq:scaling-gel})] is universal for a large variety of chemically different polymer networks and solvents within the experimental error.
Here, we used $G=E/3$ measured at the equilibrium swollen state to evaluate $\Pi_\mathrm{mix}=G$ at the equilibrium concentration $c_\mathrm{eq}$ in previous studies.
Notably, we define $c_\mathrm{eq}$ of previous studies as the prepolymer weight divided by the solvent volume, as used in this study.
The main panel in Figure~\ref{fig:previous} demonstrates the universality of the semidilute scaling law by the horizontal shift of the polymer mass concentration ($c$ to $\tilde{c}=(A'/A)^{(3\nu-1)/3\nu}c$).
Here, we denote $\Pi_\mathrm{mix}=Ac^{3\nu/(3\nu-1)}$ for this study and $\Pi_\mathrm{mix}=A'c^{3\nu/(3\nu-1)}$ for each previous study, as shown in the inset in Figure~\ref{fig:previous}.

\begin{figure}[t!]
\centering
\includegraphics[width=\linewidth]{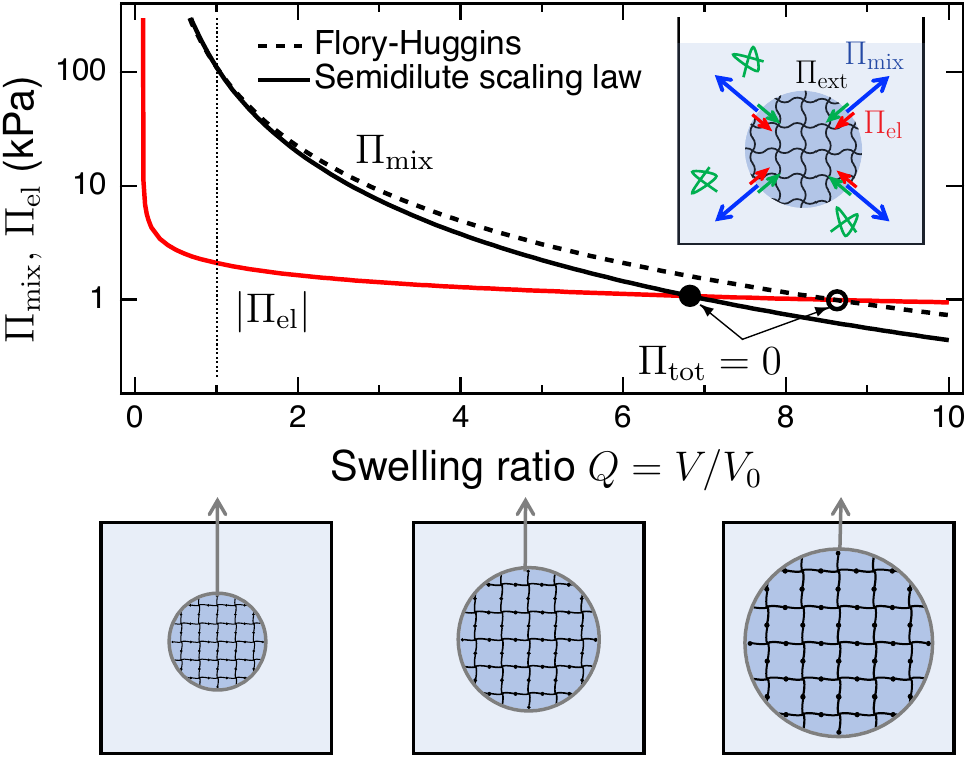}
\caption{
Decrease in $\Pi_\mathrm{mix}$ and $|\Pi_\mathrm{el}|=G_{0}(c/c_{0})^{1/3}$ (red curve) associated with swelling.
We obtained $\Pi_\mathrm{mix}$ under the FH mean-field theory (black dashed curve) and semidilute scaling law (black solid curve) from Eqs.~(\ref{eq:flory-huggins}) and (\ref{eq:scaling-gel}), respectively.
Here, we used $c$ and $\phi$ obtained from Eqs.~(\ref{eq:concentration}) and (\ref{eq:phi}).
We set the as-prepared conditions of $c_{0}$ and $G_{0}$ identical to those for the blue circles in Fig.~\ref{fig:swelling}.
The FH interaction parameter $\chi \simeq 0.439$ was adjusted at the as-prepared state ($Q=1$), which is the constant value during swelling.
The intersection points (black filled and unfilled circles) of $\Pi_\mathrm{mix}$ and $|\Pi_\mathrm{el}|$ indicate the predictions of $Q$ at the equilibrium swollen state in a pure solvent ($\Pi_\mathrm{tot}=0$).
}
\label{fig:schematic}
\end{figure}

\newpage

\section{Comparison of the two equilibrium swelling equations}
\label{App:equations}

We theoretically compare two equations [Eq.~(\ref{eq:flory-rehner})] using the FH mean-field theory [Eq.~(\ref{eq:flory-huggins})] and semidilute scaling law [Eq.~(\ref{eq:scaling-gel})] for $\Pi_\mathrm{mix}$,
to show that a small difference in $\Pi_\mathrm{mix}$ leads to a large difference in predicting the equilibrium volume swelling ratio $Q_\mathrm{eq}$ in the equilibrium swollen state.
We obtain two equilibrium equations with a certain $\Pi_\mathrm{ext}$ as follows:
\begin{equation}
nk_BTK
\left(\frac{n}{n^{*}_\mathrm{seg}}\right)
^{\frac{1}{3\nu-1}}
-G_{0}
\left(
\frac{c}{c_{0}}
\right)^{1/3}
=\Pi_\mathrm{ext},
\label{eq:Q-scaling}
\end{equation}
using the semidilute scaling law for $\Pi_\mathrm{mix}$, where $K=1.1$ and $n=cN_A/M_\mathrm{seg}$ with $M_\mathrm{seg}=44\times10^{-3}$ kg$/$mol for the ethylene glycol unit, and
\begin{equation}
-\frac{kT}{v}
\left[\phi+\ln\left(1-\phi\right)+\chi\phi^2\right]
-G_{0}
\left(
\frac{c}{c_{0}}
\right)^{1/3}
=\Pi_\mathrm{ext},
\label{eq:Q-FH}
\end{equation}
using the FH theory for $\Pi_\mathrm{mix}$, where $v=18\times10^{-6}\,\mathrm{m^{3}/mol}$ for aqueous solvents.
In equations~(\ref{eq:Q-scaling}) and (\ref{eq:Q-FH}), the volume swelling ratio $Q$ can be predicted using the experimentally measured $G_{0}$ and Eqs.~(\ref{eq:swelling-ratio}), (\ref{eq:concentration}) and (\ref{eq:phi}) for $Q$, $c$ and $\phi$.\\

Figure~\ref{fig:schematic} shows the typical change in $\Pi_\mathrm{mix}$ under the Flory--Huggins (FH) mean-field theory (black dashed curve), $\Pi_\mathrm{mix}$ under the semidilute scaling law (black solid curve), and $|\Pi_\mathrm{el}|=G_{0}(c/c_{0})^{1/3}$ (red solid curve) throughout the swelling process.
As $Q$ increases, $\Pi_\mathrm{mix}$ of the FH mean-field theory becomes larger than that of the semidilute scaling law.
The intersection points of $\Pi_\mathrm{mix}$ and $|\Pi_\mathrm{el}|$ indicate the prediction of $Q$ under the equilibrium swollen state in a pure solvent ($\Pi_\mathrm{tot}=0$).
The FH mean-field theory predicts higher $Q$ than the semidilute scaling law.

\newpage

\bibliographystyle{apsrev}

\begin{thebibliography}{50}
\expandafter\ifx\csname url\endcsname\relax
  \def\url#1{\texttt{#1}}\fi
\expandafter\ifx\csname urlprefix\endcsname\relax\def\urlprefix{URL }\fi
\providecommand{\bibinfo}[2]{#2}
\providecommand{\eprint}[2][]{\url{#2}}

\bibitem{flory1953principles}
\bibinfo{author}{P.~J.~Flory,}
\newblock \textit{\bibinfo{title}{{Principles of Polymer Chemistry}}}
 (\bibinfo{publisher}{Cornell University Press}, \bibinfo{address}{Ithaca}, \bibinfo{year}{1953}).
%Floryの教科書

\bibitem{de1979scaling}
\bibinfo{author}{P.~G.~de~Gennes,}
\newblock \textit{\bibinfo{title}{{Scaling Concepts in Polymer Physics}}}
 (\bibinfo{publisher}{Cornell University Press}, \bibinfo{address}{Ithaca}, \bibinfo{year}{1979}).
%スケーリング則

\bibitem{krishnamurti1929mechanism}
\bibinfo{author}{K.~Krishnamurti,}
\bibinfo{title}\textit{Mechanism of the swelling of gels.}
\newblock \bibinfo{journal}{Nature}
  \textbf{\bibinfo{volume}{123}}, \bibinfo{pages}{242--243}
  (\bibinfo{year}{1929}).
%ゲルの膨潤は古くから重要

\bibitem{flory1943jr}
\bibinfo{author}{P.~J.~Flory and J.~Rehner,}
\bibinfo{title}\textit{Statistial mechanics of cross-linked polymer networks II. Swelling.}
\newblock \bibinfo{journal}{J. Chem. Phys.}
  \textbf{\bibinfo{volume}{11}}, \bibinfo{pages}{521--526}
  (\bibinfo{year}{1943}).
%Flory-Rehner

\bibitem{neuburger1988critical}
\bibinfo{author}{N.~A.~Neuburger and B.~E.~Eichinger,}
\bibinfo{title}\textit{Critical Experimental Test of the Flory-Rehner Theory of Swelling.}
\newblock \bibinfo{journal}{Macromolecules}
  \textbf{\bibinfo{volume}{21}}, \bibinfo{pages}{3060--3070}
  (\bibinfo{year}{1988}).
%Flory-Rehnerで膨潤を説明/χの依存性を使用

\bibitem{wu2004modeling}
\bibinfo{author}{S.~Wu, H.~Li, J.~P.~Chen, and K.~Y.~Lam,}
\bibinfo{title}\textit{Modeling Investigation of Hydrogel Volume Transition.}
\newblock \bibinfo{journal}{Macromol. Theory Simul.}
    \textbf{\bibinfo{volume}{13}}, \bibinfo{pages}{13--29}
  (\bibinfo{year}{2004}).
%Flory-Rehnerで膨潤を説明/χの依存性を使用

\bibitem{akalp2015determination}
\bibinfo{author}{U.~Akalp, S.~Chu, S.~C.~Skaalure, S.~J.~Bryant, A.~Doostan, and F.~J.~Vernerey,}
\bibinfo{title}\textit{Determination of the polymer--solvent interaction parameter for PEG hydrogels in water: Application of a self learning algorithm.}
\newblock \bibinfo{journal}{Polymer}
    \textbf{\bibinfo{volume}{66}}, \bibinfo{pages}{135--147}
  (\bibinfo{year}{2015}).
%Flory-Rehnerで膨潤を説明/χの依存性を使用

\bibitem{patel1992elastic}
\bibinfo{author}{S.~K.~Patel, S.~Malone, C.~Cohen, and J.~R.~Gillmor,}
\bibinfo{title}\textit{Elastic Modulus and Equilibrium Swelling of Poly(dimethylsiloxane) Networks.}
\newblock \bibinfo{journal}{Macromolecules}
    \textbf{\bibinfo{volume}{25}}, \bibinfo{pages}{5241--5251}
  (\bibinfo{year}{1992}).
%Flory-Rehnerで膨潤を説明/χの依存性を使用/c*定理の否定

\bibitem{munch1977inelastic}
\bibinfo{author}{J.~P.~Munch, S.~Candau, J.~Herz, and G.~Hild,}
\bibinfo{title}\textit{Inelastic light scattering by gel modes in semi-dilute polymer solutions and permanent networks at equilibrium swollen state.}
\newblock \bibinfo{journal}{J. Phys.}
  \textbf{\bibinfo{volume}{38}}, \bibinfo{pages}{971--976}
  (\bibinfo{year}{1977}).
%c*定理の肯定：E~c^{2.31}

\bibitem{beltzung1983investigation}
\bibinfo{author}{M.~Beltzung, J.~Herz, and C.~Picot,}
\bibinfo{title}\textit{Investigation by small-angle neutron scattering of the chain conformation in equilibrium-swollen poly (dimethylsiloxane) networks.}
\newblock \bibinfo{journal}{Macromolecules}
  \textbf{\bibinfo{volume}{16}}, \bibinfo{pages}{580--584}
  (\bibinfo{year}{1983}).
%c*定理の肯定

\bibitem{bastide1984small}
\bibinfo{author}{J.~Bastide, R.~Duplessix, C.~Picot, and S.~Candau,}
\bibinfo{title}\textit{Small-Angle Neutron Scattering and Light Spectroscopy Investigation of Polystyrene Gels under Osmotic Deswelling.}
\newblock \bibinfo{journal}{Macromolecules}
  \textbf{\bibinfo{volume}{17}}, \bibinfo{pages}{83--93}
  (\bibinfo{year}{1984}).
%Flory-Rehnerで膨潤を説明/Semidilute scalingを使用

\bibitem{obukhov1994network}
\bibinfo{author}{S.~P.~Obukhov, M.~Rubinstein, and R.~H.~Colby,}
\bibinfo{title}\textit{Network modulus and superelasticity.}
\newblock \bibinfo{journal}{Macromolecules}
  \textbf{\bibinfo{volume}{27}}, \bibinfo{pages}{3191--3198}
  (\bibinfo{year}{1994}).
%c*定理の否定

\bibitem{urayama1996elastic}
\bibinfo{author}{K.~Urayama, T.~Kawamura, and S.~Kohjiya,}
\bibinfo{title}\textit{Elastic modulus and equilibrium swelling of networks crosslinked by end-linking oligodimethylsiloxane at solution state.}
\newblock \bibinfo{journal}{J. Chem. Phys.}
  \textbf{\bibinfo{volume}{105}}, \bibinfo{pages}{4833--4840}
  (\bibinfo{year}{1996}).
%c*定理の否定

\bibitem{urayama1996crossover}
\bibinfo{author}{K.~Urayama and S.~Kohjiya,}
\bibinfo{title}\textit{Crossover of the concentration dependence of swelling and elastic properties for polysiloxane networks crosslinked in solution.}
\newblock \bibinfo{journal}{J. Chem. Phys.}
  \textbf{\bibinfo{volume}{104}}, \bibinfo{pages}{3352--3359}
  (\bibinfo{year}{1996}).
%E=3G

\bibitem{candau1981experimental}
\bibinfo{author}{S.~Candau, A.~Peters, and J.~Herz}
\bibinfo{title}\textit{Experimental evidence for trapped chain entanglements: Their influence on macroscopic behaviour of networks.}
\newblock \bibinfo{journal}{Polymer}
  \textbf{\bibinfo{volume}{22}}, \bibinfo{pages}{1504}
  (\bibinfo{year}{1981}).
%E=3G

\bibitem{geissler1988compressional}
\bibinfo{author}{E.~Geissler, A.~M.~Hecht, F.~Horkay, and M.~Zrinyis,}
\bibinfo{title}\textit{Compressional modulus of swollen polyacrylamide networks.}
\newblock \bibinfo{journal}{Macromolecules}
  \textbf{\bibinfo{volume}{21}}, \bibinfo{pages}{2594--2599}
  (\bibinfo{year}{1988}).
%Flory-Rehnerで膨潤を説明/Semidilute scalingを使用/1/3のベキ

\bibitem{horkay1989effects}
\bibinfo{author}{F.~Horkay, A.~M.~Hecht, and E.~Geissler,}
\bibinfo{title}\textit{The effects of cross-linking on the equation of state of a polymer solution.}
\newblock \bibinfo{journal}{J. Chem. Phys.}
  \textbf{\bibinfo{volume}{91}}, \bibinfo{pages}{2706--2711}
  (\bibinfo{year}{1989}).
%Flory-Rehnerで膨潤を説明/Semidilute scalingを使用/1/3のベキ

\bibitem{rubinstein1996elastic}
\bibinfo{author}{M.~Rubinstein, R.~H.~Colby, and A.~V.~Dobrynin,}
\bibinfo{title}\textit{Elastic modulus and equilibrium swelling of polyelectrolyte gels.}
\newblock \bibinfo{journal}{Macromolecules}
  \textbf{\bibinfo{volume}{29}}, \bibinfo{pages}{398--406}
  (\bibinfo{year}{1996}).
%Flory-Rehnerで膨潤を説明/Semidilute scalingを使用

\bibitem{lang2014swelling}
\bibinfo{author}{M.~Lang, J.~Fischer, M.~Werner, and J.~U.~Sommer,}
\bibinfo{title}\textit{Swelling of Olympic gels.}
\newblock \bibinfo{journal}{Phys. Rev. Lett.}
  \textbf{\bibinfo{volume}{112}}, \bibinfo{pages}{238001}
  (\bibinfo{year}{2014}).
%Flory-Rehnerで膨潤を説明/Semidilute scalingを使用

\bibitem{van2015biopolymer}
\bibinfo{author}{R.~G.~M.~van~der~Sman,}
\bibinfo{title}\textit{Biopolymer gel swelling analysed with scaling laws and Flory-Rehner theory.}
\newblock \bibinfo{journal}{Food Hydrocolloids}
  \textbf{\bibinfo{volume}{48}}, \bibinfo{pages}{94--101}
  (\bibinfo{year}{2015}).
%ネットワークの膨潤：肉：Flory-Rehnerで膨潤を説明/Semidilute scalingを使用

\bibitem{horkay2000osmotic}
\bibinfo{author}{F.~Horkay, I.~Tasaki, and P.~J.~Basser,}
\bibinfo{title}\textit{Osmotic Swelling of Polyacrylate Hydrogels in Physiological Salt Solutions.}
\newblock \bibinfo{journal}{Biomacromolecules}
  \textbf{\bibinfo{volume}{1}}, \bibinfo{pages}{84--90}
  (\bibinfo{year}{2000}).
%Flory-Rehnerで膨潤を説明

\bibitem{peppas1997hydrogels}
\bibinfo{author}{N.~A.~Peppas,}
\bibinfo{title}\textit{Hydrogels and drug delivery.}
\newblock \bibinfo{journal}{Curr. Opin. Colloid Interface Sci.}
  \textbf{\bibinfo{volume}{2}}, \bibinfo{pages}{531--537}
  (\bibinfo{year}{1997}).
%ゲルの応用：ドラッグデリバリー

\bibitem{porchet1998clinical}
\bibinfo{author}{N.~de~Tribolet, F.~Porchet, T.~W.~Lutz, O.~Gratzl, J.~Brotchi, H.~A.~van~Alphen, R.~E.~van~Acker, A.~Benini, K.~N.~Strommer, R.~L.~Bernays, J.~Goffin, E.~A.~Beuls, and J.~S.~Ross,}
\bibinfo{title}\textit{Clinical assessment of a novel antiadhesion barrier gel: prospective, randomized, multicenter, clinical trial of ADCON-L to inhibit postoperative peridural fibrosis and related symptoms after lumbar discectomy.}
\newblock \bibinfo{journal}{Am. J. Orthod.}
  \textbf{\bibinfo{volume}{27}}, \bibinfo{pages}{111--120}
  (\bibinfo{year}{1998}).
%ゲルの応用：癒着防止バリア

\bibitem{hayashi2017fast}
\bibinfo{author}{K.~Hayashi, F.~Okamoto, S.~Hoshi, T.~Katashima, C.~Z.~Denise, X.~Li, M.~Shibayama, P.~G.~Elliot, U.~Chung, S.~Ohba, T.~Oshika, and T.~Sakai,}
\bibinfo{title}\textit{Fast-forming hydrogel with ultralow polymeric content as an artificial vitreous body.}
\newblock \bibinfo{journal}{Nat. Biomed. Eng.}
  \textbf{\bibinfo{volume}{1}}, \bibinfo{pages}{1--7}
  (\bibinfo{year}{2017}).
%ゲルの応用：人工硝子体

\bibitem{james1949simple}
\bibinfo{author}{H.~M.~James and E.~Guth,}
\bibinfo{title}\textit{Simple Presentation of Network Theory of Rubber, with a Discussion of Other Theories.}
\newblock \bibinfo{journal}{J. Polym. Sci.}
  \textbf{\bibinfo{volume}{4}}, \bibinfo{pages}{153--182}
  (\bibinfo{year}{1949}).
%Phantom network model

\bibitem{flory1942thermodynamics}
\bibinfo{author}{P.~J.~Flory,}
\bibinfo{title}\textit{Thermodynamics of High Polymer Solutions.}
\newblock \bibinfo{journal}{J. Chem. Phys.}
  \textbf{\bibinfo{volume}{10}}, \bibinfo{pages}{51--61}
  (\bibinfo{year}{1942}).
%Flory-Huggins

\bibitem{huggins1942some}
\bibinfo{author}{M.~Huggins,}
\bibinfo{title}\textit{Some properties of solutions of long-chain compounds.}
\newblock \bibinfo{journal}{J. Phys. Chem.}
  \textbf{\bibinfo{volume}{46}}, \bibinfo{pages}{151--158}
  (\bibinfo{year}{1942}).
%Flory-Huggins

\bibitem{huggins1942theory}
\bibinfo{author}{M.~Huggins,}
\bibinfo{title}\textit{Theory of Solutions of High Polymers.}
\newblock \bibinfo{journal}{J. Am. Chem. Soc.}
  \textbf{\bibinfo{volume}{64}}, \bibinfo{pages}{1712--1719}
  (\bibinfo{year}{1942}).
%Flory-Huggins

\bibitem{patterson1967thermodynamics}
\bibinfo{author}{D.~Patterson,}
\bibinfo{title}\textit{Thermodynamics of non-dilute polymer solutions}
\newblock \bibinfo{journal}{Rubber Chem. Technol.}
  \textbf{\bibinfo{volume}{40}}, \bibinfo{pages}{1--35}
  (\bibinfo{year}{1967}).
%Flory-Hugginsは濃厚溶液

\bibitem{merrill1993partitioning}
\bibinfo{author}{E.~W.~Merrill, K.~A.~Dennison, and C.~Sung,}
\bibinfo{title}\textit{Partitioning and diffusion of solutes in hydrogels of poly(ethylene oxide).}
\newblock \bibinfo{journal}{Biomaterials}
  \textbf{\bibinfo{volume}{14}}, \bibinfo{pages}{1117--1126}
  (\bibinfo{year}{1993}).
%PEG-水のχ

\bibitem{pedersen2005temperature}
\bibinfo{author}{J.~S.~Pedersen and C.~Sommer,}
\bibinfo{title}\textit{Temperature dependence of the virial coefficients and the chi parameter in semi-dilute solutions of PEG.}
\newblock \bibinfo{journal}{Prog. Colloid Polym. Chem.}
  \textbf{\bibinfo{volume}{130}}, \bibinfo{pages}{70--78}
  (\bibinfo{year}{2005}).
%PEG-水のχ

\bibitem{venohr1998static}
\bibinfo{author}{H.~Venohr, V.~Fraaije, H.~Strunk, and W.~Borchard,}
\bibinfo{title}\textit{Static and dynamic light scattering from aqueous poly(ethylene oxide) solutions.}
\newblock \bibinfo{journal}{Eur. Polym. J.}
  \textbf{\bibinfo{volume}{34}}, \bibinfo{pages}{723--732}
  (\bibinfo{year}{1998}).
%PEG-水のχ

\bibitem{strazielle1968light}
\bibinfo{author}{C.~Strazielle,}
\bibinfo{title}\textit{Light Diffusion Study of Heterogeneities in Polyoxyethylene Solution.}
\newblock \bibinfo{journal}{Makromol. Chem.}
  \textbf{\bibinfo{volume}{119}}, \bibinfo{pages}{50--63}
  (\bibinfo{year}{1968}).
%PEG-水のχ

\bibitem{sakai2008design}
\bibinfo{author}{T.~Sakai, T.~Matsunaga, Y.~Yamamoto, C.~Ito, R.~Yoshida, S.~Suzuki, N.~Sasaki, M.~Shibayama, and U.~Chung,}
\bibinfo{title}\textit{Design and fabrication of a high-strength hydrogel with ideally homogeneous network structure from tetrahedron-like macromonomers.}
\newblock \bibinfo{journal}{Macromolecules}
  \textbf{\bibinfo{volume}{41}}, \bibinfo{pages}{5379--5384}
  (\bibinfo{year}{2008}).
%Tetra-PEG Gel

\bibitem{boyer1945deswelling}
\bibinfo{author}{R.~F.~Boyer,}
\bibinfo{title}\textit{Deswelling of gels by high polymer solutions.}
\newblock \bibinfo{journal}{J. Chem. Phys}
  \textbf{\bibinfo{volume}{13}}, \bibinfo{pages}{363--372}
  (\bibinfo{year}{1945}).
%Osmotic deswelling

\bibitem{bastide1981osmotic}
\bibinfo{author}{J.~Bastide, S.~Candau, and L.~Leibler,}
\bibinfo{title}\textit{Osmotic Deswelling of Gels by Polymer Solutions.}
\newblock \bibinfo{journal}{Macromolecules}
  \textbf{\bibinfo{volume}{14}}, \bibinfo{pages}{719--726}
  (\bibinfo{year}{1981}).
%Osmotic deswelling

\bibitem{des1975lagrangian}
\bibinfo{author}{J.~Des~Cloizeaux,}
\bibinfo{title}\textit{The Lagrangian theory of polymer solutions at intermediate concentrations.}
\newblock \bibinfo{journal}{J. Phys. (Paris)}
  \textbf{\bibinfo{volume}{36}}, \bibinfo{pages}{281--291}
  (\bibinfo{year}{1975}).
%semidilute scaling law

\bibitem{guida1998critical}
\bibinfo{author}{R.~Guida and J.~Zinn-Justin,}
\bibinfo{title}\textit{Critical exponents of the $N$-vector model.}
\newblock \bibinfo{journal}{J. Phys. A: Math. Gen.}
  \textbf{\bibinfo{volume}{31}}, \bibinfo{pages}{8103}
  (\bibinfo{year}{1998}).
%critical exponent

\bibitem{pelissetto2002critical}
\bibinfo{author}{A.~Pelissetto and E.~Vicari,}
\bibinfo{title}\textit{Critical phenomena and renormalization-group theory.}
\newblock \bibinfo{journal}{Phys. Rep.}
  \textbf{\bibinfo{volume}{368}}, \bibinfo{pages}{549--727}
  (\bibinfo{year}{2002}).
%O(n)-symmetric

\bibitem{clisby2016high}
\bibinfo{author}{N.~Clisby and B.~D{\"u}nweg,}
\bibinfo{title}\textit{High-precision estimate of the hydrodynamic radius for self-avoiding walks.}
\newblock \bibinfo{journal}{Phys. Rev. E}
  \textbf{\bibinfo{volume}{94}}, \bibinfo{pages}{052102}
  (\bibinfo{year}{2016}).
%モンテカルロシミュレーションによる$\nu$の計算

\bibitem{kompaniets2017minimally}
\bibinfo{author}{M.~V.~Kompaniets and E.~Panzer,}
\bibinfo{title}\textit{Minimally subtracted six-loop renormalization of $O(n)$-symmetric $\phi^{4}$ theory and critical exponents.}
\newblock \bibinfo{journal}{Phys. Rev. D.}
  \textbf{\bibinfo{volume}{96}}, \bibinfo{pages}{036016}
  (\bibinfo{year}{2017}).
%$\epsilon$展開による$\nu$の計算

\bibitem{shimada2016conformal}
\bibinfo{author}{H.~Shimada and S.~Hikami,}
\bibinfo{title}\textit{Fractal dimensions of self- avoiding walks and Ising high-temperature graphs in 3D conformal bootstrap.}
\newblock \bibinfo{journal}{J. Stat. Phys.}
  \textbf{\bibinfo{volume}{165}}, \bibinfo{pages}{1006}
  (\bibinfo{year}{2016}).
%conformal boot strap法による$\nu$の計算

\bibitem{vink1971precision}
\bibinfo{author}{H.~Vink,}
\bibinfo{title}\textit{Precision measurements of osmotic pressure in concentrated polymer solutions.}
\newblock \bibinfo{journal}{Eur. Polym. J.}
  \textbf{\bibinfo{volume}{7}}, \bibinfo{pages}{1411--1419}
  (\bibinfo{year}{1971}).
%PVPの浸透圧

\bibitem{yasuda2020universal}
\bibinfo{author}{T.~Yasuda, N.~Sakumichi, U.~Chung, and T.~Sakai,}
\bibinfo{title}\textit{Universal Equation of State describes Osmotic Pressure throughout Gelation Process.}
\newblock \bibinfo{journal}{Phys. Rev. Lett.}
  \textbf{\bibinfo{volume}{125}}, \bibinfo{pages}{267801}
  (\bibinfo{year}{2020}).
%ゲルのEOS

\bibitem{bawendi1988systematic}
\bibinfo{author}{M.~G.~Bawendi and K.~F.~Freed,}
\bibinfo{title}\textit{Systematic corrections to Flory--Huggins theory : Polymer-solvent-void systems and binary blend-void systems.}
\newblock \bibinfo{journal}{J. Chem. Phys.}
  \textbf{\bibinfo{volume}{88}}, \bibinfo{pages}{2741--2756}
  (\bibinfo{year}{1988}).
%χは依存しないはず

\bibitem{pesci1989lattice}
\bibinfo{author}{A.~I.~Pesci and K.~F.~Freed,}
\bibinfo{title}\textit{Lattice theory of polymer blends and liquid mixtures: Beyond the Flory-Huggins approximation.}
\newblock \bibinfo{journal}{J. Chem. Phys.}
  \textbf{\bibinfo{volume}{90}}, \bibinfo{pages}{2017--2034}
  (\bibinfo{year}{1989}).
%χは依存しないはず

\bibitem{noda1981thermodynamic}
\bibinfo{author}{I.~Noda, N.~Kato, T.~Kitano, and M.~Nagasawa,}
\bibinfo{title}\textit{Thermodynamic Properties of Moderately Concentrated Solutions of Linear Polymers.}
\newblock \bibinfo{journal}{Macromolecules}
  \textbf{\bibinfo{volume}{14}}, \bibinfo{pages}{668--676}
  (\bibinfo{year}{1981}).
%高分子溶液のEOS

\bibitem{ohta1982conformation}
\bibinfo{author}{T.~Ohta and Y.~Oono,}
\bibinfo{title}\textit{Conformation space renormalization theory of semidilute polymer solutions.}
\newblock \bibinfo{journal}{Phys. Lett. A}
  \textbf{\bibinfo{volume}{89A}}, \bibinfo{pages}{460--464}
  (\bibinfo{year}{1982}).
%高分子溶液のEOS

\bibitem{higo1983osmotic}
\bibinfo{author}{Y.~Higo, N.~Ueno, and I.~Noda,}
\bibinfo{title}\textit{Osmotic pressure of semidilute solutions of branched polymers.}
\newblock \bibinfo{journal}{Polym. J.}
  \textbf{\bibinfo{volume}{15}}, \bibinfo{pages}{367--375}
  (\bibinfo{year}{1983}).
%高分子溶液のEOS

\bibitem{sakai2016sol}
\bibinfo{author}{T.~Sakai, T.~Katashima, T.~Matsushita, and U.~Chung,}
\bibinfo{title}\textit{Sol-gel transition behavior near critical concentration and connectivity.}
\newblock \bibinfo{journal}{Polym. J.}
  \textbf{\bibinfo{volume}{48}}, \bibinfo{pages}{629--634}
  (\bibinfo{year}{2016}).
%p=2s

\bibitem{yoshikawa2019connectivity}
\bibinfo{author}{Y.~Yoshikawa, N.~Sakumichi, U.~Chung, and T.~Sakai,}
\bibinfo{title}\textit{Connectivity dependence of gelation and elasticity in AB-type polymerization: An experimental comparison of the dynamic process and stoichiometrically imbalanced mixing.}
\newblock \bibinfo{journal}{Soft Matter}
  \textbf{\bibinfo{volume}{15}}, \bibinfo{pages}{5017--5025}
  (\bibinfo{year}{2019}).
%p=2s

\bibitem{yoshikawa2021negative}
\bibinfo{author}{Y.~Yoshikawa, N.~Sakumichi, U.~Chung, and T.~Sakai,}
\bibinfo{title}\textit{Negative Energy Elasticity in a Rubberlike Gel.}
\newblock \bibinfo{journal}{Phys. Rev. X}
  \textbf{\bibinfo{volume}{11}}, \bibinfo{pages}{011045}
  (\bibinfo{year}{2021}).
%Negative rubber elasticity

\bibitem{Sakumichi2021} N.~Sakumichi, Y.~Yoshikawa, and T.~Sakai, \textit{Linear elasticity of polymer gels in terms of negative energy elasticity}, Polym. J. \textbf{53}, 1293 (2021).

\bibitem{fujiyabu2021temperature}
\bibinfo{author}{T.~Fujiyabu, T.~Sakai, R.~Kudo, T.~Yoshikawa, T.~Katashima, U.~Chung, and N.~Sakumichi,}
\bibinfo{title}\textit{Temperature Dependence of Polymer Network Diffusion}
\newblock \bibinfo{journal}{Phys. Rev. Lett.}
 \textbf{\bibinfo{volume}{127}}, \bibinfo{pages}{237801}
 (\bibinfo{year}{2021}).
%ゲルのD

\bibitem{Shirai2022} N.~C.~Shirai and N.~Sakumichi, \textit{Solvent-Induced Negative Energetic Elasticity in Lattice Polymer Chain}, arXiv:2202.12483.

\bibitem{devanand1991asymptotic}
\bibinfo{author}{K.~Devanand and J.~C.~Selser,}
\bibinfo{title}\textit{Asymptotic behavior and long-range interactions in aqueous solutions of poly (ethylene oxide).}
\newblock \bibinfo{journal}{Macromolecules}
  \textbf{\bibinfo{volume}{24}}, \bibinfo{pages}{5943--5947}
  (\bibinfo{year}{1991}).
%$\nu=0.583(3)$

\bibitem{yamamoto1971more}
\bibinfo{author}{A.~Yamamoto, M.~Fujii, G.~Tanaka, and H.~Yamakawa,}
\bibinfo{title}\textit{More on the analysis of dilute solution data: polystyrenes prepared anionically in tetrahydrofuran.}
\newblock \bibinfo{journal}{Polym. J.}
  \textbf{\bibinfo{volume}{2}}, \bibinfo{pages}{799--811}
  (\bibinfo{year}{1971}).
%$2\nu=1.17$

\bibitem{fukuda1974solution}
\bibinfo{author}{M.~Fukuda, M.~Fukutomi, Y.~Kato, and T.~Hashimoto,}
\bibinfo{title}\textit{Solution properties of high molecular weight polystyrene.}
\newblock \bibinfo{journal}{J. Polym. Sci., Polym. Phys. Ed.}
  \textbf{\bibinfo{volume}{12}}, \bibinfo{pages}{871--890}
  (\bibinfo{year}{1974}).
%$2\nu=1.16$

\bibitem{miyaki1978excluded}
\bibinfo{author}{Y.~Miyaki, Y.~Einaga, and H.~Fujita,}
\bibinfo{title}\textit{Excluded-volume effects in dilute polymer solutions. 7. Very high molecular weight polystyrene in benzene and cyclohexane.}
\newblock \bibinfo{journal}{Macromolecules}
  \textbf{\bibinfo{volume}{11}}, \bibinfo{pages}{1180--1186}
  (\bibinfo{year}{1978}).
%$2\nu=1.19(1)$


\bibitem{ricka1984swelling}
\bibinfo{author}{J.~Ricka and T.~Tanaka,}
\bibinfo{title}\textit{Swelling of ionic gels: quantitative performance of the Donnan theory.}
\newblock \bibinfo{journal}{Macromolecules}
  \textbf{\bibinfo{volume}{17}}, \bibinfo{pages}{2916--2921}
  (\bibinfo{year}{1984}).
%charged gel

\bibitem{tang2020swelling}
\bibinfo{author}{J.~Tang, T.~Katashima, X.~Li, Y.~Mitsukami, Y.~Yokoyama, N.~Sakumichi, U.~Chung, M.~Shibayama, and T.~Sakai,}
\bibinfo{title}\textit{Swelling behaviors of hydrogels with alternating neutral/highly charged sequences.}
\newblock \bibinfo{journal}{Macromolecules}
  \textbf{\bibinfo{volume}{53}}, \bibinfo{pages}{8244--8254}
  (\bibinfo{year}{2020}).
%charged gel

\bibitem{jia2021theory}
\bibinfo{author}{D.~Jia and M.~Muthukumar,}
\bibinfo{title}\textit{Theory of Charged Gels: Swelling, Elasticity, and Dynamics.}
\newblock \bibinfo{journal}{Gels}
  \textbf{\bibinfo{volume}{7}}, \bibinfo{pages}{49}
  (\bibinfo{year}{2021}).
%charged gel

\bibitem{xinming2008polymeric}
\bibinfo{author}{L.~Xinming, C.~Yingde, A.~W.~Lloyd, S.~V.~Mikhalovsky, S.~R.~Sandeman, C.~A.~Howel, and L.~Liewen,}
\bibinfo{title}\textit{Polymeric hydrogels for novel contact lens-based ophthalmic drug delivery systems: A review.}
\newblock \bibinfo{journal}{Cont. Lens. Anterior. Eye.}
  \textbf{\bibinfo{volume}{31}}, \bibinfo{pages}{57--64}
  (\bibinfo{year}{2008}).
%ゲルの応用：コンタクトレンズ

\bibitem{karacan2007swelling}
\bibinfo{author}{C.~Karacan,}
\bibinfo{title}\textit{Swelling-induced volumetric strains internal to a stressed coal associated with CO2 sorption.}
\newblock \bibinfo{journal}{Int. J. Coal Geol.}
  \textbf{\bibinfo{volume}{72}}, \bibinfo{pages}{209--220}
  (\bibinfo{year}{2007}).
%ネットワークの膨潤：石炭

\bibitem{makitra2011determination}
\bibinfo{author}{R.~G.~Makitra, G.~G.~Midyana, and E.~Y.~Pal'chikova,}
\bibinfo{title}\textit{Determination of the interaction parameter $\chi$ in the Flory-Rehner equation for coal swelling processes.}
\newblock \bibinfo{journal}{Solid Fuel Chem.}
  \textbf{\bibinfo{volume}{45}}, \bibinfo{pages}{430--433}
  (\bibinfo{year}{2011}).
%ネットワークの膨潤：石炭

\bibitem{nelson2020mathematical}
\bibinfo{author}{H.~Nelson, S.~Deyo, S.~Granzier-Nakajima, P.~Puente, K.~Tully, and J.~Webb,}
\bibinfo{title}\textit{A mathematical model for meat cooking.}
\newblock \bibinfo{journal}{Eur. Phys. J. Plus}
  \textbf{\bibinfo{volume}{135}}, \bibinfo{pages}{1--19}
  (\bibinfo{year}{2020}).
%ネットワークの膨潤：肉

\end{thebibliography}

\end{document}